\documentclass[journal,draftclsnofoot,onecolumn,a4paper,english]{IEEEtran}
\usepackage[T1]{fontenc}
\usepackage[latin1]{inputenc}
\usepackage{theorem}
\usepackage{graphicx}
\usepackage{babel}
\usepackage{subfigure}
\usepackage{amsmath}
\usepackage{amsfonts}
\usepackage[dvips]{color}

\newcounter{example}

{\theorembodyfont{\rmfamily}
   \newtheorem{The}{{\textbf Theorem}}[section]}
{\theorembodyfont{\rmfamily}
   }
{\theorembodyfont{\rmfamily}
   \newtheorem{Lem}{{\textbf Lemma}}[section]}
{\theorembodyfont{\rmfamily}
   }

\newenvironment{example}
{\vspace{3mm}\refstepcounter{example} \noindent{\tt
Example~\theexample}\small\\} {\hfill$\diamond$\par\vskip1mm}

\begin{document}

\title{On a Family of Circulant Matrices for Quasi-Cyclic Low-Density Generator Matrix Codes}
\author{Marco~Baldi,~\IEEEmembership{Member,~IEEE,} Federico~Bambozzi, and Franco~Chiaraluce,~\IEEEmembership{Member,~IEEE}
\thanks{Copyright (c) 2011 IEEE. Personal use of this material is permitted. However, permission to use this material for any other purposes must be obtained from the IEEE by sending a request to pubs-permissions@ieee.org.

The material in this paper was presented in part at the International Symposium on Information Theory and its Applications, Auckland, New Zealand, December 2008, and
at the IEEE Information Theory Workshop, Taormina, Italy, October 2009.

M. Baldi and F. Chiaraluce are with Dipartimento di Ingegneria Biomedica, Elettronica e Telecomunicazioni,
Università Politecnica delle Marche, Ancona, Italy
(e-mail: \{m.baldi; f.chiaraluce\}@univpm.it).
F. Bambozzi is with Dipartimento di Matematica Pura e Applicata, Università degli Studi di Padova, Padova, Italy 
(e-mail: fbamb@math.unipd.it).
}}

\maketitle
\begin{abstract}
We present a new class of sparse and easily invertible circulant matrices that can have a sparse inverse though not being 
permutation matrices.
Their study is useful in the design of quasi-cyclic low-density generator matrix codes,
that are able to join the inner structure of quasi-cyclic codes with sparse generator matrices, so limiting the number of 
elementary operations needed for encoding.
Circulant matrices of the proposed class permit to hit both targets without resorting to identity or permutation matrices 
that may penalize the code minimum distance and often cause significant error floors.
\end{abstract}

\IEEEpeerreviewmaketitle

\section{Introduction}
\label{sec:Uno}

\IEEEPARstart{L}{ow}-Density Parity-Check (LDPC) codes are extremely efficient in regard to decoding algorithms, based on the message passing principle, that exploit the sparse nature of their parity-check matrices to achieve excellent performance with low complexity \cite{Richardson2001}, \cite{Wiechman2007}.

On the other hand, the generator matrix $\mathbf{G}$ of an LDPC code is usually dense and, when it is used for encoding, this gives a complexity that is quadratic in the block length.
To reduce the encoding complexity, several solutions have been proposed in the past.
Among them, some techniques aim at exploiting the sparse nature of the parity-check matrix $\mathbf{H}$ also in the encoding stage.
This can be easily achieved when $\mathbf{H}$ admits a sparse representation in lower triangular form.
When this does not occur, an {\it Approximate Lower Triangular} (ALT) version of $\mathbf{H}$ could be obtained by performing only 
row and column permutations \cite{Richardson2001EfficientEncoding}.
Alternatively, the ALT form of the parity-check matrix can be ensured by a proper design \cite{Freundlich2007}.

A different technique for low complexity encoding of LDPC codes is represented by iterative encoding \cite{Haley2002}. According to such an approach, the parity bits corresponding to each information vector are considered as erasures, and recovered by means of the message passing decoder for channels with erasures. In order for iterative encoding to be successful, the nodes associated to the parity bits must not contain a stopping set; so the structure of the parity-check matrix must be constrained. For this reason, the design of iterative encodable codes with good performance could represent a challenge \cite{Haley2005}.

One of the most effective approaches for reducing the encoding complexity is given by Low-Density Generator Matrix (LDGM) codes \cite{Cheng1996}. For such codes, $\mathbf{G}$ is also sparse and this permits to reduce significantly the amount of processing required at the encoder. 

In \cite{Garcia-Frias2003}, LDGM codes with a very sparse generator matrix were considered, and their performance estimated. It was verified that an LDGM code with Tanner graph containing degree-$1$ variable nodes exhibits high error floors. In the same paper and in \cite{Gonzalez-Lopez2007}, it was demonstrated that these floors can be substantially reduced by serially concatenating two (or more) of these codes, at the cost of an increased complexity.
In the concatenated scheme, the component codes could be selected in such a way to allow the usage of the same decoder structure for both of them \cite{Kim2006}, but serial concatenation still has consequences on complexity and latency.

LDGM codes are a wide family of codes, including, for example, concatenated single parity-check codes \cite{Oenning2001}. They also provide the core for Repeat Accumulate (RA) codes, that can be seen as the serial concatenation of an outer LDGM code and an inner accumulator \cite{Hsu2005}.
RA codes represent an alternative solution to the usage of two serially concatenated LDGM codes for reducing the error floor. 
Variants of RA codes, as Irregular RA (IRA) codes \cite{Jin2000} and Accumulate-RA (ARA) codes \cite{Abbasfar2004}, can improve performance of RA codes, and it has been proved they can also be capacity-achieving codes \cite{Pfister2007}.

Recently, an increasing interest has been devoted to Quasi-Cyclic Low-Density Parity-Check (QC-LDPC) codes, whose parity-check and generator matrices are formed by circulant blocks. Such structure of the matrices allows the usage of very simple encoding circuits, based on shift registers, that exploit the quasi-cyclic nature of the codes \cite{Li2006}.
A widespread family of QC-LDPC matrices are formed by circulant permutation blocks \cite{Fossorier2004}. Codes having this form have also been included in the amendment for mobility of the IEEE 802.16 standard \cite{802.16e}. 
Several encoding schemes are suggested in the standard; due to the almost lower triangular form of the matrices, the solution in \cite{Richardson2001EfficientEncoding} has a nearly-linear complexity. The number of operations required can be further reduced by suitable processing \cite{Yoon2005}.


So, even for QC-LDPC codes, a common approach to exploit sparse matrices for encoding is to find parity-check matrices having an almost lower triangular form \cite{802.16e}, \cite{Myung2005}. In fact, the quasi-cyclic property facilitates the hardware implementation of the encoder, but complexity (in terms of number of elementary operations) still depends on the density of the matrix used for encoding. 
However, to find almost lower triangular parity-check matrices is not possible in several cases.
Moreover, when the parity-check matrix of a QC-LDPC code is formed by circulant blocks that are not permutation matrices,
the corresponding generator matrix is usually dense.
This occurs when the row (column) weight of each non-null circulant block in the parity-check matrix is greater than one,
and can be found, for example, in the QC-LDPC codes proposed for near-Earth missions by the Consultative Committee for Space Data Systems (CCSDS) \cite{Li2006}, \cite{CCSDS2007}.
Another interesting family of QC-LDPC codes having circulant blocks with row (column) weight greater than one are those based on Difference Families and their variants \cite{Johnson2003, Xia2005}. The structure of the parity-check matrix of these codes is as follows:
\begin{equation}
\mathbf{H} = [\mathbf{H}_0 | \mathbf{H}_1 | ... | \mathbf{H}_{N_b-1}],
\label{eq:HCircRow}
\end{equation}
i.e., it consists of a row of $N_b$ sparse circulant blocks, each with size $n=N/N_b$, where $N$ is the code length. 
Provided that at least one of the $\mathbf{H}_i$ blocks is of full rank, the code rate is $(N_b-1)/N_b$.
Despite their very simple structure, codes having this form can be able to achieve good performance, especially for moderate/high code rates.

For codes having a parity-check matrix in the form (\ref{eq:HCircRow}), a low-density generator matrix can be found if one of the $\mathbf{H}_i$ blocks is replaced with an $n \times n$ identity matrix or cyclic permutation matrix. This way, however, the minimum distance of the code is penalized, and becomes less than or equal to the lowest row (column) weight of the non-identity $\mathbf{H}_i$ blocks, increased by $1$.

In this paper, we define a new class of sparse circulant matrices, that we call $\psi$-unitary (the reason for such notation will be explained afterwards). These matrices are simple to design, easily invertible and can have a sparse inverse, though not being circulant permutation matrices. Furthermore, they can be free of length-$4$ cycles; so, we propose to use them for constructing the parity-check matrix in the form (\ref{eq:HCircRow}). 
By replacing one of the $\mathbf{H}_i$ blocks with a $\psi$-unitary matrix, the density of the code generator matrix can be rendered very low while maintaining a good minimum distance.

The features of the new class of matrices are derived by extending the theory of orthogonal circulant matrices (\cite{MacWilliams71}, \cite{Zhang1997phd}), that are a special case of circulant matrices but that cannot be used for the design of QC-LDPC codes. This is explained in Section \ref{sec:Due}, which is devoted to remind basic definitions and properties. The $\psi$-unitary circulant matrices are introduced in Section \ref{sec:DueB}, where conditions for the absence of length-$4$ cycles are explicitly stated. Section \ref{sec:Tre} presents some families of matrices that are free of length-$4$ cycles, while Section \ref{sec:TreB} discusses the inversion issues. 
The inversion method we propose has complexity that depends mainly on the matrix weight and is basically independent of the matrix size; moreover, availability of explicit expressions for the inverse matrix permits us to estimate its density. In Section \ref{sec:Quattro}, examples of usage of $\psi$-unitary circulant matrices in LDGM codes are given. Finally, Section \ref{sec:Cinque} concludes the paper.

\section{Circulant matrices: notation and properties}
\label{sec:Due}

The general structure of an $n \times n$ circulant matrix $\mathbf{A}$ defined over the Galois field
of order $p$, $GF(p)$, is as follows:
\begin{equation}
\mathbf{A}=\left[\begin{array}{ccccc}
a_{0} & a_{1} & a_{2} & \cdots & a_{n-1}\\
a_{n-1} & a_{0} & a_{1} & \cdots & a_{n-2}\\
\vdots & \vdots & \vdots & \ddots & \vdots\\
a_{1} & a_{2} & a_{3} & \cdots & a_{0}\end{array}\right],
\label{eq:Circulant}
\end{equation}
where $a_i \in GF(p)$, $i = 0 \ldots n-1$. 
Thus, $\mathbf{A}$ is described by one of its rows (typically the first one),
the others being obtained as cyclically shifted versions of such row.
In the following, we will denote by $W[\mathbf{A}]$ the number of non-zero symbols in each row (or column) of $\mathbf{A}$.

A simple isomorphism exists between the ring $M_n$ of $n \times n$ circulant matrices over 
$GF(p)$ and the ring $R_n = GF(p)[x]/(x^n-1)$ of the polynomials over $GF(p)$ modulo $(x^n - 1)$. 
Let us consider the following $n \times n$ circulant permutation matrix:
\begin{equation}
\mathbf{T}=\left[\begin{array}{ccccc}
0 & 1 & 0 & \cdots & 0\\
0 & 0 & 1 & \cdots & 0\\
\vdots & \vdots & \vdots & \ddots & \vdots\\
0 & 0 & 0 & \cdots & 1\\
1 & 0 & 0 & \cdots & 0\end{array}\right].\label{eq:MatrixT}\end{equation}
It it easy to verify that the circulant matrix (\ref{eq:Circulant}) can be written as:
\begin{equation}
\mathbf{A}=\sum_{i=0}^{n-1}a_i\mathbf{T}^i,
\label{eq:Omo1}
\end{equation}
where $\mathbf{T}^0 = \mathbf{T}^n = \mathbf{I}$. 
This relationship is the basis to establish the isomorphism, defined by the following map:
\begin{equation}
\phi:\sum_{i=0}^{n-1}a_i\mathbf{T}^i \rightarrow \sum_{i=0}^{n-1}a_ix^i = a(x),
\label{eq:Omo2}
\end{equation}
that transforms matrices into polynomials modulo $(x^n - 1)$. 
The minimal polynomial of $\mathbf{T}$ is $\mathbf{T}^n-\mathbf{I}$. 
According to this isomorphism, we can work, when more convenient, with polynomials instead of matrices. 
So, from now on, a matrix $\mathbf{A}$ will be equivalently denoted by the polynomial $a(x)$; 
$W[a]$ is the weight of $a(x)$ (number of its non-zero coefficients) and it coincides with $W[\mathbf{A}]$. 
According to (\ref{eq:Omo2}), the polynomial $a(x)$ is specified by its coefficients $(a_0, a_1, ..., a_{n-1})$.

A length-$4$ cycle in matrix $\mathbf{A}$ is a closed rectangular path linking non-zero elements.
Explicitly, this means that a length-$4$ cycle exists when two rows have a pair of non-zero symbols at the same positions (i.e., belonging to the same columns). Obviously, each matrix can have manifold loops of this kind, depending on the distribution of the non-zero symbols.
It is easy to verify that length-$4$ cycles in matrix $\mathbf{A}$
do not appear if and only if the distances between any pair of non-zero symbols in each row of the matrix are different from each other (explicitly, we say that the matrix has no repeated distances).
So, if $\delta_{i,k}$ represents the distance (mod $n$) between the $i$-th and $k$-th non-zero elements, it must be $\delta_{i,k} = \delta_{j,l}$ if and only if $i = j$ and $k = l$.

Orthogonal circulant matrices are a special case of circulant matrices, for which $\mathbf{A} \cdot \mathbf{A}^T = \mathbf{I}$. So, the inverse of an orthogonal matrix coincides with its transpose.
The following theorem holds:
\begin{The} \label{Th:DF1}
An orthogonal circulant matrix $\mathbf{A}$ over $GF(p)$, with $W[\mathbf{A}] > 1$, has always length-$4$ cycles.

\begin{IEEEproof}
Let us suppose that matrix (\ref{eq:Circulant}) is orthogonal; then, the inner product between its $i$-th and $j$-th rows must be 0, for $i \neq j$.  Without loss of generality, let us consider $i=0$; the following condition must be satisfied:
\[a_0a_{n-j}+a_1a_{n-j+1}+...+a_va_m+...+a_{n-1}a_{n-j-1}= 0,\]
where all subscripts are mod $n$. If a column of $\mathbf{A}$ has at least two non-zero elements (which means it is not a permutation matrix, the latter being a particular case, not of interest for the present analysis), say $a_v$ and $a_m$, with $m=(v-j)$ mod $n$, then $a_va_m \neq 0$. Condition above implies there exists at least another term $a_wa_y \neq 0$, with $a_w$ belonging to the same row of $a_v$, and $a_y$ to the same row of $a_m$, and $y = (w-j)$ mod $n$. Therefore, $a_v$, $a_m$, $a_w$ and $a_y$ define a length-$4$ cycle, since $v - m = w - y = j$ mod $n$.
\end{IEEEproof}
\end{The}
Because of Theorem \ref{Th:DF1}, orthogonal circulant matrices with weight greater than 1 are not suitable for the design of QC-LDPC codes. However, starting from the theory developed in previous literature (see \cite{MacWilliams71} and \cite{Zhang1997phd}) for the study and characterization of orthogonal circulant matrices, it is possible to define a new class of matrices that, instead, can be free of length-$4$ cycles. This is done in the following section.

\section{$\psi$-unitary circulant matrices}
\label{sec:DueB}

\subsection{Definition and properties}
\label{sec:DueB1}
Let us consider $n=ps$, with $s$ an integer, and the ring $R_s = GF(p)[y]/(y^s-1)$. Let us define the following map from the ring $R_n$ to the ring $R_s$:
\begin{equation}
\psi_s^{ps}: \sum_{i = 0}^{n-1} a_i x^i \to \sum_{k = 0}^{s-1} u_k y^k,
\label{eq:mappsi1}
\end{equation}
where
\begin{equation}
u_k = \sum_{t = 0}^{p-1} a_{k+st}, \:\: \forall k \in \left\{0, 1, \ldots, s-1\right\}.
\label{eq:Sist}
\end{equation}
The map transforms elements of $R_n$ into elements of $R_s$, according to the specified rule.

\begin{example}
Let us consider $\psi_2^4: R_4 \rightarrow R_2$ over $GF(2)$. The elements of $R_4$ are:
\begin{eqnarray*}
(0,0,0,0), (0,0,0,1), (0,0,1,0), (0,0,1,1), \\
(0,1,0,0), (0,1,0,1), (0,1,1,0), (0,1,1,1), \\
(1,0,0,0), (1,0,0,1), (1,0,1,0), (1,0,1,1), \\
(1,1,0,0), (1,1,0,1), (1,1,1,0), (1,1,1,1).
\end{eqnarray*}
So:
\begin{equation*}
\begin{split}
\psi_2^4((0,0,0,0)) & = \psi_2^4((0,1,0,1)) = \psi_2^4((1,0,1,0)) = \\
& = \psi_2^4((1,1,1,1)) = (0,0),  \\
\psi_2^4((1,0,0,0)) & = \psi_2^4((0,0,1,0)) = \psi_2^4((1,1,0,1)) = \\
& = \psi_2^4((0,1,1,1)) = (1,0),  \\
\psi_2^4((0,1,0,0)) & = \psi_2^4((0,0,0,1)) = \psi_2^4((1,1,1,0)) = \\
& = \psi_2^4((1,0,1,1)) = (0,1),  \\
\psi_2^4((1,1,0,0)) & = \psi_2^4((0,1,1,0)) = \psi_2^4((1,0,0,1)) = \\
& = \psi_2^4((0,0,1,1)) = (1,1).
\end{split}
\end{equation*}
The same elements can also be written in polynomial form:
\begin{equation*}
\begin{split}
& \psi_2^4(0) = \psi_2^4(x+x^3) = \psi_2^4(1+x^2) = \psi_2^4(1 +x+x^2+x^3) = 0, \\
& \psi_2^4(1) = \psi_2^4(x^2) = \psi_2^4(1+x+x^3) = \psi_2^4(x+x^2+x^3) = 1, \\
& \psi_2^4(x) = \psi_2^4(x^3) = \psi_2^4(1+x+x^2) = \psi_2^4(1+x^2+x^3) = x, \\
& \psi_2^4(1+x) = \psi_2^4(x+x^2) = \psi_2^4(1+x^3) = \psi_2^4(x^2+x^3) = 1+x.
\end{split}
\end{equation*}
\end{example}

It is known \cite{MacWilliams71} that $\psi_s^{ps}$ is an homomorphism, with kernel $Ker(\psi_s^{ps}) = I_s^{ps}$, where $I_s^{ps} = \left< x^s - 1 \right> \in R_n$ denotes the ideal generated by $(x^s - 1)$. $I_s^{ps}$ is completely described by the following formula:
\begin{equation}
a(x) = \sum _{i =0}^{n - 1} a_i x^i  \in I_s^{ps}  \Leftrightarrow \sum_{t=0}^{p-1} a_{k + ts}= 0,
\label{DefIn}
\end{equation}
for all $k \in \left\{0, 1, \ldots, s-1\right\}$.

Generalizing (\ref{eq:mappsi1}), we can consider $n=p^q s$ and the map $\psi^{p^q s}_{p^r s}: R_{p^q s} \to R_{p^r s}$, with $0 \leq r \leq q$ and $Ker(\psi^{p^q s}_{p^r s}) = I^{p^q s}_{p^r s} = \left< x^{p^r s} - 1 \right>$ (if $r = q$, $\psi^{p^q s}_{p^q s}$ is the identity map). From a computational point of view, the map $\psi^{p^q s}_{p^r s}$ corresponds to iterating ($q-r$) times the elementary mapping from $R_{p^t s}$ to $R_{p^{t-1} s}$, with $q \geq t > r$. Noting by $\psi(c(x))$ the result of the single mapping, applied to the polynomial $c(x)$, we should write:
\begin{eqnarray*}
\psi^{p^q s}_{p^r s}(a(x)) = \psi^{p^{r+1}s}_{p^r s}(\psi^{p^{r+2}s}_{p^{r+1}s}(...\psi^{p^qs}_{p^{q-1}s}(a(x))...)).
\end{eqnarray*}

Let us denote by $1_{p^r s}$ the identity element of $R_{p^r s}$. The following definition holds:

{\it Definition:} A circulant matrix in $M_{p^q s}$ is a $\psi$-unitary circulant matrix if
its polynomial $a(x) \in R_{p^q s}$, for some $r \leq q$, satisfies the condition:
\begin{equation}
\label{definition}
\psi^{p^q s}_{p^r s}(a(x)) = 1_{p^r s}.
\end{equation}

$\psi$-unitary matrices, so defined, form a subset of the matrices in $M_{p^q s}$, that in the following will be denoted by $\Psi^{p^q s}_{p^r s}$. It is evident that, if a matrix $\mathbf{A} \in \Psi^{p^q s}_{p^r s}$ for a given $r$, then it satisfies the condition $\psi^{p^q s}_{p^t s}(a(x)) = 1_{p^t s}$ for any $0 \leq t < r$.

\begin{example}
\label{Ex:1}
Let us consider $p = 2$, $r = 0$, $q = 1$ and $s = 4$ (hence, $n = 2s = 8$). The following polynomials with weight 3 
satisfy the condition $\psi^{8}_{4} = 1_{4} = (1,0,0,0)$; so, they define $\psi$-unitary circulant matrices of size $8 \times 8$:
$(a_0,a_1,a_2,a_3,a_4,a_5,a_6,a_7)=
(1,1,0,0,0,1,0,0)$,  
$(1,0,0,1,0,0,0,1)$, 
$(0,1,0,0,1,1,0,0)$,
$(0,0,0,1,1,0,0,1)$,
$(1,0,1,0,0,0,1,0)$,
$(0,0,1,0,1,0,1,0)$.
\end{example}

For the sake of simplicity, and because this work aims at designing binary codes, from now on we will set $p = 2$. Moreover, as we are mainly interested in considering sparse matrices, it is often convenient to denote $a(x)$ by the positions $b_j$ of its $W[a]$ non-zero elements ($1 \leq j \leq W[a]$), that is $a(x)=(b_1; b_2; ... ; b_{W[a]})_n$.

\begin{figure}[t]
\centering
\includegraphics[keepaspectratio,scale=0.6]{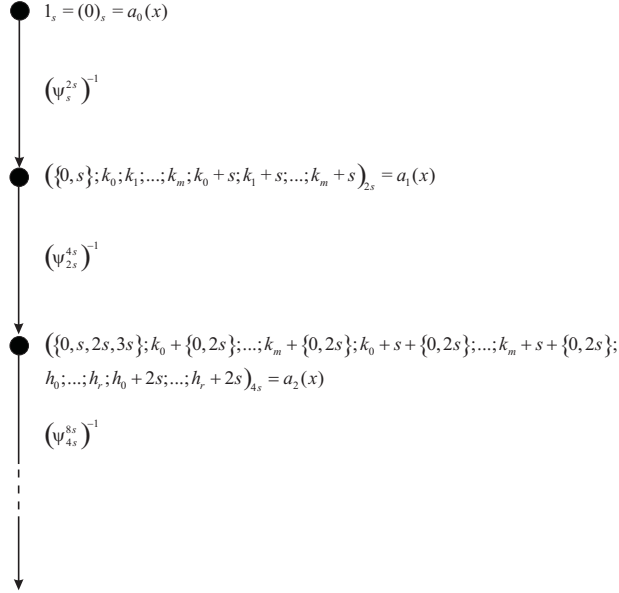}
\caption{Inverse mapping representation for a $\psi$-unitary circulant matrix ($\{a, b, ..., c\}$ means that all the listed options are possible).} 
\label{fig:HomInv}
\end{figure}

For the subsequent analysis, it will be essential to consider the inverse mapping $\psi^{-1}$. This is schematically shown in Fig. \ref{fig:HomInv}, where the case of a $\psi$-unitary circulant matrix is considered.
Basically, the inverse mapping $(\psi^{2^q s}_{2^{q-1} s}(a(x)))^{-1}$ adds to $a(x)$ an element of the ideal $I^{2^q s}_{2^{q-1} s}$. In the following, this will be indicated as $\{a(x)+I^{2^q s}_{2^{q-1} s}\}$.
Moreover, because of (\ref{DefIn}), for any $w(x) \in I^{2^q s}_{2^{q-1} s}$, it is easy to verify that $[w(x)]^2=0$, and vice versa.
So, if we define the map $\eta: R_n \to R_n$ as $\eta[a(x)] = [a(x)]^2$, it follows that $w(x) \in I^{2^q s}_{2^{q-1} s} \Leftrightarrow w(x) \in Ker(\eta)$. 
By iterating the reasoning, and denoting $r$ consecutive applications of $\eta$ as $\eta^r$, it follows that:
\begin{equation}
Ker(\psi_{2^{q-r}s}^{2^q s}) = Ker(\eta^r).
\label{eq:KernelEquivalence}
\end{equation}

This property can be used to demonstrate the equivalence between (\ref{definition}) and a condition on the powers of $a(x)$. For simplifying the notation, in the following we will set: $[a(x)]^z = a^z(x)$. So, in particular, $a^{-z}(x)=[a^{-1}(x)]^z$ will denote the $z$-power of the polynomial $a^{-1}(x)$, that corresponds to the inverse matrix $\mathbf{A}^{-1}$.
The following lemma holds:
\begin{Lem}
\label{Lem1}
For $n = 2^q s$ and $a(x) \in R_n$, $a^{2^r}(x) = 1_{2^q s} \Leftrightarrow \psi^{2^q s}_{2^{q - r} s}(a(x)) = 1_{2^{q - r} s}$, with $0 < r \leq q$.
%

\begin{IEEEproof}
If $\psi^{2^q s}_{2^{q - r} s}(a(x)) = 1_{2^{q - r} s}$, $a(x)$ must be in the form $a(x) = b(x) + w(x)$, where $b(x)$ is a monomial mapped into $1_{2^{q-r} s}$ by $\psi^{2^q s}_{2^{q - r} s}$ and 
$w(x) \in I^{2^q s}_{2^{q-r} s}$. It follows from its definition that $b(x) = (j 2^{q-r} s)_n$, $j \in \left\{0, 1, \ldots, 2^r-1\right\}$, so $b^{2^r}(x) = (0)_n = 1_n$.
Due to \eqref{eq:KernelEquivalence}, $w^{2^r}(x) = 0$, so $a^{2^r}(x) = b^{2^r}(x) + w^{2^r}(x) = 1_n$.
The reverse implication follows by the same argument.
\end{IEEEproof}
\end{Lem}

The coincidence established by Lemma \ref{Lem1} between polynomials $a(x)$ such that $a^{2^r}(x) = 1_{2^q s}$ and $\psi^{2^q s}_{2^{q - r} s}(a(x)) = 1_{2^{q - r}s}$
will be recalled in the following (as in the proofs of Theorems \ref{The222} and \ref{The232}).

\subsection{Existence of length-$4$ cycles}
\label{sec:DueB2}
Let us consider $q = 1$ and $r = 0$; the $\psi$-unitary matrices, defined by (\ref{definition}), form the subset $\Psi^{2s}_{s}$. From (\ref{eq:Sist}) we derive that, in order to have $u_0 = 1$, that is necessary to have $\mathbf{A} \in \Psi^{2s}_{s}$,
either $a_0$ or $a_s$ must be different from zero. Based on this evidence, we can formulate the following:
\begin{The}
\label{The216}
For $n = 2s$, a matrix $\mathbf{A} \in \Psi^{2s}_{s}$, with $W[\mathbf{A}]>1$, has length-$4$ cycles.

\begin{IEEEproof}
Taking into account (\ref{eq:mappsi1}) and (\ref{eq:Sist}), to have $\mathbf{A} \in \Psi^{2s}_{s}$, $W[\mathbf{A}]$ should be necessarily odd. Hence, the polynomial $a(x)$ must have an odd number of non-zero coefficients. Starting from $a_0 = 1$ or $a_s = 1$, the matrix weight becomes $W[\mathbf{A}] = 3$ by adding, for any $k: 0 < k < s$, a non-zero coefficient $a_k$ and a non-zero coefficient $a_{k+s}$; therefore, the distance $n/2$  between $a_k$ and $a_{k+s}$ is equal to the distance between $a_{k+s}$ and $a_k$ (repeated distance) and, as explained in Section \ref{sec:Due}, this is the reason for the appearance of a length-$4$ cycle between any row of the matrix and its $s$-shifted version. Such repeated distance is always present but, depending on the value of $k$, other repeated distances can appear. Similarly, for $W[\mathbf{A}] > 3$, the additional non-zero coefficients are always at distance $n/2$, and therefore contribute to increase the multiplicity of this repeated distance.
\end{IEEEproof} 
\end{The}

\begin{example}
\label{Ex:2}
The matrices generated by the polynomials in the Example \ref{Ex:1} have length-$4$ cycles. In fact, considering the non-zero elements, we can verify that the  polynomial $(1,1,0,0,0,1,0,0)$,  
$(1,0,0,1,0,0,0,1)$, 
$(0,1,0,0,1,1,0,0)$ and
$(0,0,0,1,1,0,0,1)$ have $1$ repeated distance ($n/2 = 4$), while the polynomials
$(1,0,1,0,0,0,1,0)$ and
$(0,0,1,0,1,0,1,0)$ have $3$ repeated distances ($2$, $4$ and $6$).
\end{example}

In order to find matrices free of length-$4$ cycles, the following theorem is useful:
\begin{The}
\label{The223}
For $n = 2^qs$, with $q > 1$, a matrix $\mathbf{A} \in \Psi^{2^qs}_{s}$ can be free of length-$4$ cycles for $W[\mathbf{A}] \leq 2q-1$; on the other hand, matrix $\mathbf{A}$ has length-$4$ cycles for $W[\mathbf{A}]>2q-1$.

\begin{IEEEproof}
Let us consider at first the case $q=2$, that is $n = 4s$. By applying $\psi^{4s}_{2s}$ to the polynomial $a(x)$ we obtain a polynomial $a'(x)$ that defines a matrix $\mathbf{A}'$. Obviously, we have $W[\mathbf{A}] \geq W[\mathbf{A}']$. So, when $W[\mathbf{A}] = 3$, it must be $W[\mathbf{A}'] \leq 3$ and odd. In the case $W[\mathbf{A}'] = 1$, matrix $\mathbf{A}$ certainly has length-$4$ cycles. 
This is because Theorem \ref{The216} can be applied.

In the case $W[\mathbf{A}'] = 3$, instead, it must be $a'(x)= (0; k; k+s)_{2s}$ or $a'(x)= (s; k; k+s)_{2s}$. By considering the first case (demonstration is similar for the second one), we have $a(x)= (\{0,2s\}; k+\{0,2s\}; k+s+\{0,2s\})_{4s}$; by examining the structure of these polynomials, it is easy to verify that, except for some values of $k$, in relation with the value of $s$, the corresponding matrices do not exhibit repeated distances, and therefore they are free of length-$4$ cycles.

When $W[\mathbf{A}] > 3$, matrix $\mathbf{A}$ has always length-$4$ cycles. For example, when $W[\mathbf{A}] = 5$, three cases are possible:
\begin{enumerate}
\item $W[\mathbf{A}'] = 5$; then $a(x)= (\{0,s, 2s, 3s\}; k_0+\{0,2s\}; k_1+\{0,2s\}; k_0+s+\{0,2s\}; k_1+s+\{0,2s\})_{4s}$, and it is easy to verify that, regardless the values of $k_0$ and $k_1$, these polynomials always exhibit repeated distances.
\item $W[\mathbf{A}'] = 3$; then $a(x)= (\{0,s, 2s, 3s\}; k_0+\{0,2s\}; k_0+s+\{0,2s\}; h_0; h_0 + 2s)_{4s}$, and the repeated distance $2s = n/2$ appears, which is due to the symbols 1 at positions $h_0$ and $h_0 + 2s$.
\item $W[\mathbf{A}'] = 1$; then $a(x)= (\{0,s, 2s, 3s\}; h_0; h_1; h_0 + 2s; h_1 + 2s)_{4s}$, and there are multiple repeated distances equal to $n/2$.
\end{enumerate}
The analysis can be immediately extended to the case of $W[\mathbf{A}] > 5$, where the same conclusions hold.

Now let us consider the general case with $q > 2$. Based on the demonstration above, it is clear that a necessary condition for having $\mathbf{A}$  free of length-$4$ cycles is that $W[\mathbf{A}] = W[\mathbf{A}']$, where $\mathbf{A}'$ is the matrix relative to the polynomial $a'(x) = \psi^{2^q s}_{2^{q - 1} s}(a(x))$. In fact, in all cases where $W[\mathbf{A}] > W[\mathbf{A}']$, the additional elements in $a(x)$ must necessarily be at distance $n/2$, thus generating a repeated distance.
In order to complete the proof, we can proceed by induction from small matrices to larger ones, through the application of the inverse mapping. 
Let us suppose that $a'(x)$ defines a matrix $\mathbf{A}'$, with $n' = 2^{q - 1}s$, free of length-$4$ cycles. According to the induction hypothesis, its weight is $W[\mathbf{A}'] \leq 2(q - 1) - 1$. Let us consider this relationship with the equality sign (i.e., we assume the maximum weight). It is not difficult, by applying the inverse mapping $(\psi^{2^q s}_{2^{q-1} s}(a(x)))^{-1}$, to obtain a polynomial $a(x)$ defining a matrix $\mathbf{A}$ with the same weight of $\mathbf{A}'$ and also free of length-$4$ cycles. For example, the rightmost (or leftmost) part of $a(x)$ can be coincident with $a'(x)$ (but this is not, obviously, the only solution).

If we now consider another polynomial, $a''(x)$, defining a matrix $\mathbf{A}''$ with $n'' = 2^{q - 1}s$ but weight $W[\mathbf{A}''] = 2q - 1$, according to the induction hypothesis, surely it has repeated distances (and, therefore, length-$4$ cycles). Actually, it is not difficult to design a matrix $\mathbf{A}''$ that contains only one repeated distance. To this purpose, it suffices to start from $a'(x)$ and to add two elements at distance $n''/2 = n/4$. When applying the inverse mapping $(\psi^{2^q s}_{2^{q-1} s}(a''(x)))^{-1}$, these two elements translate into elements in $a(x)$ that are at distance $n/4$ in their turn. This demonstrates that a matrix $\mathbf{A}$ can exist, with $n = 2^qs$ and $W[\mathbf{A}] = 2q - 1$, that is free of length-$4$ cycles. An explicit rule for its construction will be given in Theorem \ref{The231}.

If, always starting from $a'(x)$, the number of added elements is four instead of two, the weight of $\mathbf{A}''$ becomes $W[\mathbf{A}''] = 2q + 1$. Because of the definition of $\psi$-unitary matrix, these further elements must be placed in positions that necessarily introduce (at least) another repeated distance. By applying the inverse homomorphism, two (or more) repeated distances in $a'(x)$ will translate into one (or more) repeated distance in $a(x)$, and matrix $\mathbf{A}$ will certainly contain length-$4$ cycles. The same reasoning obviously applies for larger weights, so that we can conclude
that a matrix $\mathbf{A}$, with $n = 2^qs$ and $W[\mathbf{A}] > 2q - 1$, has always length-$4$ cycles.

\end{IEEEproof}
\end{The}

\section{Design of $\psi$-unitary circulant matrices free of length-$4$ cycles}
\label{sec:Tre}
Theorem \ref{The223} states that $\psi$-unitary binary circulant matrices can exist, that are free of length-$4$ cycles and have an arbitrary odd weight (through a proper choice of $q$). However, it does not provide an explicit structure for these matrices. The latter can be found on the basis of other theorems, that will be given next.

We now introduce the sets of polynomials we will use to design parity-check matrices of QC-LDGM codes. To make more explicit the notation, we will often put, in the following, $q = m + 2$ and assume $n = 2^{m+2}s$. The sets of interest can be described by polynomials having the structure (\ref{eq:cosets}), 
where $m \geq 0$, $0<k_0<s, 0<k_1<2s, ... , 0<k_m<2^ms$, $k_i \not= k_j$, $\forall i \not= j$ and $c_i$, $d_i$ are integer coefficients: $0 \leq c_i, d_i < 2^{m+1-i}$:
\begin{align}
a(x) = 	& (c_{-1} s ; k_0 + c_0 2 s; k_1 + c_1 4 s ; \ldots; k_{m} + c_m 2^{m+1} s; \nonumber \\
				&  k_0+s + d_0 2 s; k_1 + 2s + d_1 4 s; \ldots; \nonumber \\
				&  k_{m}+ 2^{m} s+ d_m 2^{m+1}s)_{2^{m+2}s}.
\label{eq:cosets}
\end{align}


The set of matrices (\ref{eq:cosets}) will be denoted by $\Xi^{2^{m+2}s}_s$; clearly $\Xi^{2^{m+2}s}_s \subset \Psi^{2^{m+2}s}_s$.
A matrix in $\Xi^{2^{m+2}s}_s$ can be free of length-$4$ cycles under a suitable choice of the $k_i$'s (we note this is coherent with Theorem \ref{The223}). The following theorem holds:
\begin{The}
\label{The231}
Let us consider $n=2^{m+2}s$. The matrix:
\begin{equation}
a(x)=(0; k_0; k_1;...;k_m; k_0 + s; k_1 + 2s; ...; k_m + 2^ms)_n,
\label{eq:goodmat}
\end{equation}
with $0<k_0<k_1<...<k_m$, $k_{i+1}>2k_{i}$ and $s > 2k_m$, is free of length-$4$ cycles.

\begin{IEEEproof}
We note that $a(x) \in \Xi^{2^{m+2}s}_s$. Proof is immediate by calculating all possible distances between symbols 1 in $a(x)$. Because of the geometric progression, when fixing the attention on the $i$-th position, the distances $\delta_{i,j}$, with $j < i$, are all different one each other and always greater than the distances $\delta_{i-1,k}$, with $k < i-1$. Moreover, because of the assumption on the value of $n$, independently of $m$, it is always $\delta_{i,j} < \delta_{j,i}$ for $i > j$. As an example, the distance between the first and the last position is $3 \cdot 2^m s - k_m > 2^ms + k_m$, the latter being the distance between the last and the first position; similarly for the other distances. This ensures that repeated distances cannot exist in $a(x)$. 
\end{IEEEproof}
\end{The}
Since the weight of $a(x)$ in (\ref{eq:goodmat}) is $W[a] = 2m + 3$, Theorem \ref{The231} gives an explicit rule to design matrices free of length-$4$ cycles with such a weight.

\begin{example}
\label{ex:p3}
Let us set $m = 1$, $s = 7$ and $n=2^{m+2}s = 56$ and choose $k_0 = 1$ and $k_1 = 3$. The following matrix:
\[a(x) = (0; 1; 3; 8; 17)_{56}\]
is free of length-$4$ cycles.
\end{example}

Obviously, we cannot say that (\ref{eq:goodmat}) is the only structure able to ensure the absence of length-$4$ cycles. Even more, $a(x)$ could have the structure (\ref{eq:goodmat}) but without satisfying the relationships between the $k_i$'s and $s$ specified by the theorem, while remaining free of length-$4$ cycles.

\begin{example}
\label{ex:p4}
Let us consider the matrix:
\[a(x) = (0; 1; 3, 7; 12; 25; 51)_{176},\]
that has the structure (\ref{eq:goodmat}) with $m = 2, k_0 = 1, k_1 = 3, k_2 = 7$, and $s = 11 < 2k_2$. It is possible to verify, through explicit calculation, that this matrix is free of length-$4$ cycles.
\end{example}


It is important to note that the interest on the structure (\ref{eq:cosets}) is justified by its simplicity and the possibility of fast matrix inversion through the procedure described in the next section.

In Section \ref{sec:TreB} we also derive bounds on the weight of the inverse of a $\psi$-unitary matrix, and we show that its actual weight can be very low with respect to the matrix size.
Moreover, the bounds we derive on the weight of the inverse do not depend on the matrix size; so, such matrices are able to provide encoding complexity that is linear in the code length.
These are the reasons why $\psi$-unitary matrices are of interest for the
design of QC-LDPC codes in the form of LDGM codes. 

\section{$\psi$-unitary matrix inversion}
\label{sec:TreB}
A standard algorithm for inverting an $n \times n$ circulant matrix exploits the fact that any matrix of this type can be made diagonal through the so-called Fourier matrix. This permits for a very fast inversion. The main limitation of this approach is that, when operating in a ring $\mathbb{Z}_m$, inversion is possible if and only if $m$ and $n$ are coprime. In addition, it is also necessary to find an extension of the ring, in such a way as to guarantee the presence of $n$ roots in the unit circle, as required by the implementation of the fast Fourier transform. These limitations can be overcome by exploiting the isomorphism between matrices and polynomials \cite{Bini2001}.

\subsection{Explicit evaluation of the inverse matrix}
For a matrix $\mathbf{A}$ with $n = 2s$ (i.e., $q = 1$) satisfying the condition $\psi^{2s}_{s} = 1_s$, Lemma \ref{Lem1} implies $a^2(x) = 1_{2s}$, and therefore $a^{-1}(x) = a(x)$, which means that $\mathbf{A}$ coincides with its inverse.
This result permits to prove the following:

\begin{The}
\label{The222}
If $n = 4s$ (i.e., $q = 2$) and $\mathbf{A_0} \in \Psi^{4s}_{s}$ then, $\forall a(x) \in \lbrace a_0(x) + I^{4s}_{2s} \rbrace$, the following relationship holds: $a(x) + a^{-1}(x) = w(x)$, where $w(x) \in I^{4s}_{2s}$ is independent of the choice of $a(x)$.

\begin{IEEEproof}
As $\psi_s^{4s} = 1_s$, from Lemma \ref{Lem1} we have: $a_0^4(x) = 1_{4s}$. This implies $a_0^2(x) \cdot a_0^2(x) = 1_{4s}$ and then $a_0^2(x) = a_0^{-2}(x)$. By applying the map $\psi^{4s}_{2s}$ to both sides of the equality, because of the homomorphism properties, one finds $[\psi^{4s}_{2s}(a_0(x))]^2 = [\psi^{4s}_{2s}(a_0^{-1}(x))]^2$.
To verify this equality, $a_0(x)$ and $a_0^{-1}(x)$ must differ by an element belonging to the ideal $I^{4s}_{2s}$, i.e., it must be:
\[a_0(x) + \{a_0^{-1}(x) + w_0(x)\} = 0,\]
for some $w_0(x) \in I^{4s}_{2s}$. Similarly, by considering another element $a_1(x) \in \lbrace a_0(x) + I^{4s}_{2s} \rbrace \neq a_0(x)$ we shall find a $w_1(x) \in I^{4s}_{2s}$ such that:
\[a_1(x) + \{a_1^{-1}(x) + w_1(x)\} = 0.\]
So:
\[a_0(x) + a_0^{-1}(x) + w_0(x) = a_1(x) + a_1^{-1}(x) +w_1(x).\]
To demonstrate that $w_0(x) = w_1(x) = w(x)$ (that is, uniqueness of $w(x)$) it is sufficient to prove that:
\[a_0(x) + a_0^{-1}(x) = a_1(x) + a_1^{-1}(x).\]
For this purpose, we observe that $a_1(x) \in \lbrace a_0(x) + I^{4s}_{2s} \rbrace$ can be always written as follows:
\[a_1(x) = a_0(x) + w_2(x).\]
Then, the above equality is verified if and only if:
\[a_1^{-1}(x) = a_0(x) + w_0(x) + w_2(x).\]
But this relationship is certainly true. In fact, by using it, we have, as necessary:
\begin{eqnarray*}
a_1(x)a_1^{-1}(x) & = & \left[a_0(x) + w_2(x) \right] \left[a_0(x) + w_0(x) + w_2(x) \right] \\
									& = & a_0^2(x) + a_0(x)w_0(x) \\
									& = & a_0(x)a_0^{-1}(x) = 1_{4s}
\end{eqnarray*}
having exploited the fact that $w_2^2(x) = 0$ and $w_0(x)w_2(x) = 0$ (this is because, by definition, the non-zero elements $w(x) \in I^{4s}_{2s}$ are of type $b(x)(x^{2s} + 1)$, where $b(x)$ is a polynomial with the maximum order $< 2s$).
\end{IEEEproof}
\end{The}

For the matrices satisfying the hypotheses of Theorem \ref{The222}, the inverse matrix can be found by adding $w(x)$ to $a(x)$; $w(x)$ can be calculated by inverting one matrix of the set $\lbrace a_0(x) + I^{4s}_{2s} \rbrace$, and using it for all matrices of the set.

Theorem \ref{The222} can be extended as follows:
\begin{The}
\label{The232}
If $n = 2^{m+2}s$ and $\mathbf{A_0} \in \Psi^{2^{m+2}s}_{s}$ then, $\forall a(x) \in \lbrace a_0(x) + I^{2^{m + 2}s}_{2^{m + 1} s} \rbrace$, the following relationship holds:
\begin{equation}
a^{-1}(x) = (a^{2^{m}}(x) + w(x)) \prod_{i = 0}^{m - 1} a^{2^{i}}(x),
\label{eq:Inverter}
\end{equation}
where $w(x) \in I^{2^{m + 2}s}_{2^{m + 1} s}$ is independent of the choice of $a(x)$.

\begin{IEEEproof}
Proceeding like in the demonstration of Theorem \ref{The222}, we can see that, by Lemma \ref{Lem1}, if $\psi^{2^{m+2}s}_{s}(a_0(x)) = 1_s$ then 
\[ a_0^{2^{m+2} }(x) = 1_{2^{m+2} s}\;\;\; \rightarrow \;\;\; a_0^{2^{m+1} }(x) = a_0^{-2^{m+1} }(x).\]
From this:
\[ [a_0^{2^{m} }(x)]^2 = [a_0^{-2^{m} }(x)]^2\]
and then:
\[ \left[\psi^{2^{m+2}s}_{2^{m+1}s}(a_0^{2^{m} }(x)) \right]^2 = \left[\psi^{2^{m+2}s}_{2^{m+1}s}(a_0^{-2^{m} }(x)) \right]^2.\]
This equality is satisfied by assuming:
\[ a_0^{-2^m}(x) = a_0^{2^m}(x) + w(x), \]
for $w(x) \in I^{2^{m+2} s}_{2^{m + 1} s}$.
The same holds for any other $a(x) \in \lbrace a_0(x) + I^{2^{m + 2}s}_{2^{m + 1} s} \rbrace$, and the polynomial $w(x)$ is unique (demonstration as in Theorem \ref{The222}).
Replacing $a_0(x)$ by $a(x)$, to obtain $a^{-1}(x)$, we need to multiply both sides of the above relationship by:
\[a^{2^{m}-1}(x) = \prod_{i = 0}^{m - 1} a^{2^{i}}(x),\]
from which (\ref{eq:Inverter}) is derived.
\end{IEEEproof}
\end{The}

\subsection{Fast inversion}
\label{F_inv}
Eq. (\ref{eq:Inverter}) provides a direct method for the computation of the inverse of a $\psi$-unitary circulant matrix, that can be much faster than more conventional methods. The key point of the procedure is the calculation of the polynomial $w(x)$. Multiplying both sides of (\ref{eq:Inverter}) by $a(x)$, we see that $w(x)$ can be obtained, in general, as the solution of the following equation:
\begin{equation}
a^{2^m}(x)w(x) = 1 + a^{2^{m+1}}(x).
\label{eq:invmod}
\end{equation}
However, for the matrices in the ensemble $\Xi^{2^{m+2}s}_s$, that has been defined in Section \ref{sec:Tre}, Eq. (\ref{eq:invmod}) can be directly solved.

More precisely, it can be verified that, for these matrices,
\begin{align}
a^{2^m}(x) = 	& 2^m(c_{-1} s; k_0 + c_0 2s; k_1; \\
  						&	k_0+s + d_0 2s; k_1+ 2s((1 + \delta_{m,0})))_{2^{m+2}s},
\label{eq:EqLem3}
\end{align}
where $\delta_{i,j}$ is the Kronecker delta function, and
\begin{equation}
a^{2^{m+1}}(x) = 	2^{m+1}(c_{-1} s; k_0 + c_0 2s; k_0+s + d_0 2s)_{2^{m+2}s}.
\label{eq:EqLem4}
\end{equation}

By replacing (\ref{eq:EqLem3}) and (\ref{eq:EqLem4}) 
in (\ref{eq:invmod}), the expression of $w(x)$ can be explicitly found and then, through (\ref{eq:Inverter}), the polynomial of the inverse, $a^{-1}(x)$, can also be obtained. 

On the basis of the previous expressions, it could seem that $c_{-1} = 0, 1, 2, 3$, $c_0 = 0,1$ and $d_0 = 0, 1$, in all their possible $16$ combinations, define as many different situations. Actually, it is possible to verify, even through explicit calculation, that the structure of $w(x)$ depends only on the value of $c_{-1}$. 

In detail, for even $c_{-1}$ (that is, $c_{-1}=0, 2$), we have:
\begin{equation}
\label{eq:Weven}
w(x)_e= 2^m(2k_0; 3k_0; 3k_0+s; 2k_0+2s; 3k_0+2s; 3k_0+3s)_n
\end{equation}
while, for odd $c_{-1}$ (that is, $c_{-1}=1, 3$), we have:
\begin{eqnarray}
\label{eq:Wodd}
w(x)_o 	& = & 2^m(k_0; 3k_0; s; k_0+s; 2k_0+s; 3k_0+s; k_0+2s; \nonumber \\
				&		& 3k_0+2s; 3s; k_0+3s; 2k_0+3s; 3k_0+3s)_n.
\end{eqnarray}
It is interesting to note that $w(x)$ is independent of $k_1$.
It can be easily proved that $W[a^{2^m} + w] \leq 11$; this result will be useful in the following (see Appendix \ref{app:1}, in particular).

According to (\ref{eq:Weven}) and (\ref{eq:Wodd}), we see that, for computing $w(x)$, we can always refer to the case $m=0$, as the effect of $m > 0$ simply results in multiplying the polynomial so obtained by $2^m$.
The values of $c_{-1}$, $c_0$ and $d_0$ determine the structure of $a(x)$ and, eventually, that of $a^{-1}(x)$. For $m=0$, however, it is possible to verify that the following combinations: ($c_{-1}, c_0, d_0$) = ($0, 0, 0$), ($1, 1, 0$), ($2, 1, 1$) and ($3, 0, 1$) are equivalent, in the sense they define polynomials that differ by a cyclic shift of $s = n/4$ positions, or a multiple of it. Precisely, these polynomials are: $(0;k_0;s+k_0)_{4s}$, $(s;s+k_0;2s+k_0)_{4s}$, $(2s;2s+k_0;3s+k_0)_{4s}$ and $(k_0;3s;3s+k_0)_{4s}$ respectively. Similarly, ($0, 0, 1$), ($1, 0, 0$), ($2, 1, 0$) and ($3, 1, 1$) are also equivalent, and the same holds for combinations ($0, 1, 0$), ($1, 1, 1$), ($2, 0, 1$), ($3, 0, 0$) and for combinations ($0, 1, 1$), ($1, 0, 1$), ($2, 0, 0$), ($3, 1, 0$). For any equivalent set, only one matrix must be considered. In fact, if two polynomials differ by a right shift of $s$ positions, their inverses differ by a left shift of $s$ positions as well.

When required, in the following we will focus on the choice $c_{-1} = 0$, that is the first matrix of each equivalent set.

\subsection{Comparison with Euclid's algorithm}
The inverse of a matrix satisfying the assumptions of Theorem \ref{The232} can be easily found by using (\ref{eq:Inverter}). This reflects in a computation algorithm that, in many cases, can be significantly faster than more conventional approaches for matrix inversion. In this subsection, in particular, we give some examples of comparison between the proposed approach and the more classic Euclid's algorithm. 

A distinctive feature of the proposed approach, against Euclid's algorithm, is that it exhibits a much weaker dependence on the matrix size. In fact, the complexity of our approach is mainly influenced by the matrix weight. This can make the proposed procedure highly effective in the case of large and sparse matrices, like those of interest for LDPC code design.

Some numerical examples are given in Table \ref{tab:speed} for different values of $n$ and different weights $W[a]$. The table shows the average time required for a single matrix inversion using Euclid's algorithm and the new algorithm. For each considered case, the processing time values have been obtained by simulating the inversion of a suitably large set of matrices (assumed to be the same for both algorithms) randomly chosen in the ensemble $\Xi^{2^{m+2}s}_s$.

\begin{table*}[ht]
\caption{Normalized average time for matrix inversion using Euclid's algorithm and the new algorithm}

\begin{centering}
\begin{tabular}{|c|ccccccc|}
\hline
$n$ & $128$ & $256$ & $512$ & $1024$ & $2048$ & $4096$ & $8192$ \\
\hline
$W[a]=3$ &  &  &  &  &  &  & \\
\hline
Euclid's alg. & 1.00 & 1.96 & 4.19 & 9.13 & 22.88 & 60.59 & 174.39 \\
New alg. & 0.22 & 0.22 & 0.23 & 0.24 & 0.23 & 0.23 & 0.23 \\
\hline
$W[a]=5$ &  &  &  &  &  &  & \\
\hline
Euclid's alg. & 3.17 & 9.99 & 32.07 & 108.89 & 552.78 & 2145.35 & 9511.17  \\
New alg. & 0.52 & 0.55 & 0.58 & 0.59 & 0.57 & 0.61 & 0.60 \\
\hline
$W[a]=7$ &  &  &  &  &  &  & \\
\hline
Euclid's alg. & 4.33 & 13.98 & 50.11 & 196.99 & 941.40 & 4083.91 & 16769.77 \\
New alg. & 2.29 & 2.96 & 3.79 & 4.47 & 4.86 & 5.33 & 5.45 \\
\hline
$W[a]=9$ &  &  &  &  &  &  & \\
\hline
Euclid's alg. & 4.91 & 16.25 & 60.21 & 232.80 & 1069.89 & 4802.12 & 20076.41 \\
New alg. & 4.64 & 9.53 & 21.86 & 34.40 & 44.08 & 86.63 & 94.11 \\
\hline
\end{tabular}
\par\end{centering}
\label{tab:speed} 
\end{table*}

Both inversion algorithms were implemented in PARI/GP \cite{PARI2009}, by exploiting optimized libraries for polynomials over finite fields.
The source code was profiled by considering only the operation of inversion, without any accessory function, as those for loading matrices and storing results.
All simulations ran on a fixed hardware and the numerical values have been normalized with respect to the average inversion time required by Euclid's algorithm on matrices with $n = 128$ and $W[a] = 3$, that was $0.87$ ms on the hardware adopted.
Only small values of $W[a]$ have been considered.
From the table we see that the processing time required by the new algorithm is smaller than that required by Euclid's algorithm for the considered weights, that are of interest for the design of LDPC codes.


\subsection{Weight of the inverse matrix}
\label{subsec:InverseWeight}
Another important consequence of the availability of explicit expressions for $w(x)$, and then for $a^{-1}(x)$, is the possibility to estimate the weight of the inverse matrix. To this purpose, the following theorems hold:

\begin{The}
\label{The224}
For $n = 4s$, the matrices in $\Xi^{4s}_s$ with $W[a] = 3$, free of length-$4$ cycles, have $5 \leq W[a^{-1}] \leq 9$.

\begin{IEEEproof}
Let us consider the first combination of each equivalent set, that has $c_{-1}=0$. Because of Theorem \ref{The222}, by using (\ref{eq:EqLem3}) and (\ref{eq:Weven}) with $m = 0$, we obtain:
\begin{eqnarray*}
a^{-1}(x) & = & (0; k_0+c_0 2 s; 2k_0; k_0 + s + d_0 2 s; 3k_0; \\
& &  3k_0+s; 2k_0+2s; 3k_0+2s; 3k_0+3s)_{4s}.
\end{eqnarray*}
Depending on the values of $k_0$ and $s$, at most four positions can be coincident two by two; therefore, the weight of $a^{-1}(x)$ is between 5 and 9 (the latter when no positions are coincident).

\end{IEEEproof}
\end{The}

For $W[a] > 3$ the weight of the inverse is destined to increase and the evaluation of a range of variability for $W[a^{-1}]$ becomes more difficult as well. In Section \ref{sec:Quattro}, however, where the proposed theory will be applied for the design of LDGM codes, we will limit to consider $W[a] \leq 5$ (i.e., $m = 0$ and $m = 1$), as this weight is large enough to ensure the achievement of good performance.

Similarly to Theorem \ref{The224}, that holds for $m = 0$, an upper bound for $W[a^{-1}]$ in the case of $m = 1$ can be found through explicit calculation. More precisely, starting from (\ref{eq:Inverter}) and using the properties of the matrices in $\Xi^{8s}_s$ it is easy to find:
\begin{gather}
a^{-1}(x) = (a^2(x)+w(x))a(x) = \nonumber \\
(0;2k_0;4k_0;6k_0;2k_1;2s + 2k_0;4s + 2k_1;2s + 6k_0;4s + 4k_0; \nonumber \\
4s + 6k_0;6s + 6k_0;k_0;3k_0;5k_0;7k_0;2k_1 + k_0;2s + 3k_0; \nonumber \\
4s + 2k_1 + k_0;2s + 7k_0;4s + 5k_0;4s + 7k_0;6s + 7k_0;k_1; \nonumber \\
2k_0 + k_1;4k_0 + k_1;3k_1;4s + 3k_1;4s + 4k_0 + k_1;s + k_0; \nonumber\\
s + 3k_0;s + 5k_0;s + 7k_0;s + 2k_1 + k_0;3s + 3k_0; \nonumber \\
5s + 2k_1 + k_0;3s + 7k_0;5s + 5k_0;5s + 7k_0;7s + 7k_0; \nonumber \\
2s + k_1;2s + 4k_0 + k_1;2s + 3k_1;4s + 2k_0 + k_1; \nonumber \\
6s + 3k_1;6s + 4k_0 + k_1)_{8s},
\label{eq:Inv1}
\end{gather}
that provides $W[a^{-1}] \leq 45$. Obviously, this upper bound has a sense only for $s$ sufficiently large ($s \geq 6$); otherwise, the upper bound would be greater than $n$.

Depending on the values of $k_0$ and $k_1$, the upper bound can be lower. This occurs when some terms, in (\ref{eq:Inv1}), become equal and, therefore, annul each other. As an example, for $k_1 = 3k_0$ the maximum weight is $21$.

Moreover, depending on the value of $s$, the actual weight of the inverse can be smaller.

\begin{example}
Let us consider the matrix of the Example \ref{ex:p3}. By applying (\ref{eq:Weven}), we find $w(x) = (4; 6; 20; 32; 34; 48)_{56}$. Then, using (\ref{eq:Inverter}), the inverse matrix is obtained as:
\begin{eqnarray*}
a^{-1}(x) & = & (a^2(x) + w(x))a(x) \\
					& = & \left( (0;2;6;16;34)_{56} + (4;6;20;32;34;48)_{56}  \right) \cdot \\
					&		& \cdot (0;1;3;8;17)_{56} \\
					& = & (0;2;4;16;20;32;48)_{56} \cdot (0;1;3;8;17)_{56} \\
					& = & (1;2;4;7;8;9;10;12;16;20;23;24;28;32;35;37; \\
					&		& 40;48;51)_{56}.
\end{eqnarray*}
Thus $W[a^{-1}]=19$, which is smaller than the upper bound.
\end{example}

Another interesting issue concerns the distribution of the inverse matrix weight. For $m = 1$, some examples are shown in Table \ref{tab:p51a}, 
where we have reported the (per cent) incidence of each weight, estimated through a Montecarlo simulation of $100,000$ matrices of each size.
We see that, for $s > 16$, the weight spectrum has a maximum at the upper bound. The convergence to the upper bound, confirmed by the trend of the average value $\left\langle W[a^{-1}] \right\rangle $, becomes more and more evident for increasing $s$, since further elisions in the expression above become less and less probable. Explicitly, this means that, for very large $n$, the upper bound gives the actual weight for an increasing fraction of the inverse matrices.

In Appendix \ref{app:1}, a bound on the weight of the inverse is derived for the case $m > 1$.

\begin{center}
\begin{table}[ht]
\caption{Weight distribution of the inverse matrices in the case of $m = 1$, for different values of $s$}
\begin{centering}
\begin{tabular}{|c|rrrr|}
\hline
$W[a^{-1}]$ & $s=16$ & $s=64$ & $s=256$ & $s=1024$ \\ 
\hline
$5$ & $0.04$ & $0$ & $0$ & $0$ \\
$7$ & $0.08$ & $0.01$ & $0$ & $0$ \\
$9$ & $0.08$ & $0.01$ & $0$ & $0$ \\
$11$ & $0.04$ & $0.01$ & $0$ & $0$ \\
$13$ & $0.67$ & $0.03$ & $0$ & $0$ \\
$15$ & $0.93$ & $0.07$ & $0.01$ & $0$ \\
$17$ & $0.93$ & $0.06$ & $0$ & $0$ \\
$19$ & $1.04$ & $0.07$ & $0.01$ & $0$ \\
$21$ & $3.22$ & $0.91$ & $0.23$ & $0.05$ \\
$23$ & $4.51$ & $1.14$ & $0.29$ & $0.07$ \\
$25$ & $7.10$ & $1.15$ & $0.27$ & $0.06$ \\
$27$ & $4.10$ & $0.84$ & $0.17$ & $0.05$ \\
$29$ & $2.89$ & $0.63$ & $0.16$ & $0.05$ \\
$31$ & $0.46$ & $0.03$ & $0$ & $0$ \\
$33$ & $2.29$ & $0.12$ & $0.01$ & $0$ \\
$35$ & $16.77$ & $4.03$ & $1.00$ & $0.23$ \\
$37$ & $11.60$ & $4.97$ & $1.33$ & $0.34$ \\
$39$ & $10.02$ & $3.17$ & $0.81$ & $0.20$ \\
$41$ & $20.62$ & $10.80$ & $3.05$ & $0.77$ \\
$43$ & $8.99$ & $6.63$ & $1.93$ & $0.48$ \\
$45$ & $3.62$ & $65.32$ & $90.73$ & $97.70$ \\
\hline
$\left\langle W[a^{-1}] \right\rangle $ & $34.95$ & $42.40$ & $44.35$ & $44.84$ \\
\hline
\end{tabular}
\par\end{centering}
\label{tab:p51a} 
\end{table}
\par\end{center}

\section{$\psi$-unitary matrices in LDGM codes}
\label{sec:Quattro}

\subsection{Code features}
\label{subsec:LDGMintro}

The LDGM codes considered in \cite{Garcia-Frias2003} are systematic codes with generator matrix $\mathbf{G} = [\mathbf{I} | \mathbf{P}]$, where $\mathbf{P}$ is a $K \times (N - K)$ sparse matrix and $\mathbf{I}$ is the $K \times K$ identity matrix. The parity-check matrix of these codes can be expressed as $\mathbf{H} = [\mathbf{P}^T | \mathbf{I}]$, where the identity matrix has size $(N-K) \times (N-K)$ and superscript $^T$ denotes transposition. So, matrix $\mathbf{H}$ is also sparse, and this permits the application of standard algorithms for decoding of LDPC codes. 

The weakness of these codes is the existence, in the Tanner graph, of $N - K$ coded (parity) bit nodes with degree $1$. This implies that the messages propagated from the order-$1$ coded bits to their corresponding check bits are always the same and are not affected by the decoding algorithm. As a consequence, these codes exhibit high error floors, which require resorting to serial concatenation of two LDGM codes \cite{Gonzalez-Lopez2007}.

It is reasonable to think that the performance of LDGM codes, both in single and concatenated configuration, can be improved by replacing the identity matrix in $\mathbf{H}$, that is responsible for the order-$1$ coded bits, with some other sparse matrix having weight larger than $1$. So, contrary to \cite{Garcia-Frias2003}, our starting point is the parity-check matrix. 

Let us consider (\ref{eq:HCircRow}).
By assuming that at least one of the $\mathbf{H}_{i}$ blocks ($i=0 \ldots N_b-1$), e.g., the last block, is of full rank, the generator matrix $\mathbf{G}$ can be obtained as:
\begin{equation}
\mathbf{G}=\left[\mathbf{I}\begin{array}{c}
\left(\mathbf{H}_{N_{b}-1}^{-1}\cdot\mathbf{H}_{0}\right)^{T}\\
\left(\mathbf{H}_{N_{b}-1}^{-1}\cdot\mathbf{H}_{1}\right)^{T}\\
\vdots\\
\left(\mathbf{H}_{N_{b}-1}^{-1}\cdot\mathbf{H}_{N_{b}-2}\right)^{T}\end{array}\right];
\label{eq:Gsystematic}
\end{equation}
so, it is formed by a $K \times K$ identity matrix (remind that $K$ is the information length) followed by a column of $K_b = N_b - 1$ circulant blocks with size $n = N/N_b$. Matrices $\mathbf{G}$ and $\mathbf{H}$ are related through the expression $\mathbf{H} \cdot \mathbf{G}^T = \mathbf{0}$.

We observe that the LDGM codes in \cite{Garcia-Frias2003} and \cite{Gonzalez-Lopez2007} can be interpreted as a special case of the codes with (\ref{eq:HCircRow}) and (\ref{eq:Gsystematic}), where $\mathbf{H}_{N_{b}-1} = \mathbf{I}$.
The choice of the identity matrix gives the lowest possible density of the generator matrix, but at the expense of the code minimum distance. As mentioned, this reflects on high error floors.
We will denote such codes as \textit{identity} (or $\textit{I}$-based) QC-LDGM codes in the following.

The alternative choice we propose consists in using, as $\mathbf{H}_{N_{b}-1}$, a $\psi$-unitary block with suitable weight (greater than $1$). We have seen in Section \ref{sec:TreB} that $\psi$-unitary matrices, properly designed, can have sparse inverses, so producing LDGM codes in the specified sense. They ensure easy encoding and good decoding features without penalizing the distance properties.
We will denote such codes as $\psi$-unitary QC-LDGM codes in the following, and we will compare their performance and complexity with those of  
$\textit{I}$-based codes both in analytical and numerical terms.

In order to avoid the existence of length-$4$ cycles in the Tanner graph associated to each $\psi$-unitary QC-LDGM code,
we adopt $\psi$-unitary matrices in the form \eqref{eq:goodmat} as $\mathbf{H}_{N_{b}-1}$.
They are designed by randomly choosing the $k_0 \ldots k_m$ parameters, with the constraints $0<k_0<k_1<...<k_m$ and $s > 2k_m$ (see Theorem \ref{The231}). Though the further constraint $k_{i+1}>2k_{i}$ has not always been imposed, matrices free of length-$4$ cycles have been found for all the considered choices of code parameters.
The selection of $\psi$-unitary matrices has been done by aiming at reducing the weight of the inverse matrix
with respect to the bounds found in Section \ref{subsec:InverseWeight}, as much as possible.
The circulant blocks $\mathbf{H}_i$, $i \in \left\{0, 1, \ldots, N_{b}-2\right\}$, have been designed on a random basis, by avoiding
the introduction of length-$4$ cycles with the $\psi$-unitary block and between each couple of them. 

All codes we consider can be treated as LDPC codes, and decoded through standard belief propagation
algorithms. We adopt the log-likelihood ratios sum-product algorithm (LLR-SPA) \cite{Hagenauer1996}.

In order to compare performance of $\psi$-unitary QC-LDGM codes with that of $\textit{I}$-based QC-LDGM 
codes, we first refer to transmission over the Binary Symmetric Channel (BSC).
Then, we will give some examples of performance over the Additive White Gaussian Noise (AWGN) channel, where 
we will also assess the concatenated scheme proposed for $\textit{I}$-based codes \cite{Garcia-Frias2003}, 
\cite{Gonzalez-Lopez2007}.

\subsection{Minimum distance and multiplicity}
\label{subsec:MinDist}

In order to estimate the minimum distance and its multiplicity for $\textit{I}$-based and $\psi$-unitary codes,
we can refer to previous literature.
For parity-check matrices in the form \eqref{eq:HCircRow}, it is proved in \cite{Kamiya2007} 
that, $\forall i,j: 0 \leq i < j < N_b$, there exists a codeword of weight
$W[\mathbf{H}_i\mathbf{H}_j] \leq W[\mathbf{H}_i] + W[\mathbf{H}_j]$; so, the code minimum
distance can be upper bounded as follows:
\begin{equation}
d_{\min} \leq \displaystyle\min_{0 \leq i < j < N_b}\left\{W[\mathbf{H}_i] + W[\mathbf{H}_j]\right\} = \overline{d}_{\min}.
\label{eq:dminBound}
\end{equation}
Based on these arguments, we can also obtain a (loose) lower bound on the number of 
weight-$\overline{d}_{\min}$ codewords as follows:
\begin{equation}
P_{\overline{d}_{\min}} = \left| \left\{ i,j: W[\mathbf{H}_i] + W[\mathbf{H}_j] = \overline{d}_{\min}; 0 \leq i < j < N_b \right\} \right|.
\end{equation}
Each of the $P_{\overline{d}_{\min}}$ low weight codewords involves a different pair of circulant blocks $\left(i,j\right)$;
so, it cannot coincide with a cyclically shifted version of another of such codewords.
We denote each of them as a \textit{low weight pattern} in the following.
The number $P_{\overline{d}_{\min}}$ of low weight patterns can be easily estimated
starting from the two smallest block weights and their block multiplicity:
\begin{equation}
\left\{ 
\begin{array}{ccl}
W_1 & = & \displaystyle\min_{0 \leq i < N_b}{W[\mathbf{H}_i]}; \\
W_2 & = & \displaystyle\min_{0 \leq i < N_b}\left\{W[\mathbf{H}_i]: W[\mathbf{H}_i] > W_1\right\}; \\
N_1 & = & \left| \left\{ \mathbf{H}_i: W[\mathbf{H}_i] = W_1 \right\} \right|; \\
N_2 & = & \left| \left\{ \mathbf{H}_i: W[\mathbf{H}_i] = W_2 \right\} \right|.
\end{array}
\right.
\label{eq:MinBlockWeight}
\end{equation}
Based on \eqref{eq:MinBlockWeight}, $P_{\overline{d}_{\min}}$ can be estimated as follows:
\begin{equation}
P_{\overline{d}_{\min}} = 
\left\{ 
\begin{array}{ccl}
N_2, & & \mathrm{if} \ N_1 = 1; \\ 
{N_1 \choose 2}, & & \mathrm{if} \ N_1 > 1.
\end{array}
\right.
\label{eq:Pdmin}
\end{equation}
Due to the quasi-cyclic nature of the codes, each of the $P_{\overline{d}_{\min}}$ low weight
patterns can give rise, at most, to $n$ cyclically shifted versions of itself that are still valid 
codewords.
So, an estimate of the weight-$\overline{d}_{\min}$ codewords multiplicity can be expressed as:
\begin{equation}
A_{\overline{d}_{\min}} \approx n \cdot P_{\overline{d}_{\min}}.
\label{eq:AdminMult}
\end{equation}
For the codes considered in Section \ref{subsec:Examples}, we have found that \eqref{eq:dminBound} holds with
the equality sign and we have verified \eqref{eq:Pdmin} by analyzing undetected errors 
(or decoder errors) that occur due to transitions of the received codeword to near codewords
during Montecarlo simulations.
In almost all cases, we have found a number of different low weight patterns exactly coincident with that
predicted by \eqref{eq:Pdmin}.
The only exception was the $(8192, 7168)$ $\psi$-unitary code, for which only $21$
different low weight patterns (out of $28$ predicted by \eqref{eq:Pdmin}) were found. 

\subsection{Complexity assessment}
\label{subsec:Complexity}

In order to compare $\textit{I}$-based with $\psi$-unitary QC-LDGM codes under the complexity viewpoint,
we need to estimate both their encoding and decoding requirements.

An exact complexity evaluation should be referred to a specific implementation, and depends on a variety
of factors as the degree of parallelization, the routing strategies and the memory occupation.
All these aspects are influenced by the hardware architecture adopted and the design choices.
On the contrary, we need a complexity measure independent of the final implementation, but significant 
enough for a fair comparison between different codes.

For this reason, we express complexity in terms of the number of elementary operations needed for
encoding and decoding. Such number is strictly related to the density of symbol $1$ in the generator 
and parity-check matrices, and allows to compare complexity of different codes without referring to any
specific hardware or software implementation.

As a measure of encoding complexity, we consider the number of elementary operations needed to calculate
each redundancy bit. So, encoding complexity can be expressed as the average column weight in the last
$N-K$ columns of the generator matrix in systematic form:

\begin{equation}
C_{enc} = \frac{\sum_{i=K}^{N-1} W\left[\mathbf{g}_i \right]}{N-K},
\label{eq:Cenc}
\end{equation}
where $W\left[\mathbf{g}_i \right]$ denotes the Hamming weight of the $i$-th column of $\mathbf{G}$.

We consider generator matrices in the form (\ref{eq:Gsystematic}); so, the column weight in their last
$N-K$ columns is constant and coincident with the sum of the Hamming weights of a row (or column) of
$\mathbf{H}_{N_{b}-1}^{-1}\cdot\mathbf{H}_{0}, \mathbf{H}_{N_{b}-1}^{-1}\cdot\mathbf{H}_{1}, \ldots, 
\mathbf{H}_{N_{b}-1}^{-1}\cdot\mathbf{H}_{N_{b}-2}$.

For an $\textit{I}$-based QC-LDGM code, $\mathbf{H}_{N_{b}-1}$ is an identity block, and the
Hamming weights of the blocks in the non-identity part of $\mathbf{G}$ coincide with those 
of the first $N_{b}-1$ blocks of $\mathbf{H}$. If they all
have row (or column) weight $X$, it results in $C_{enc} = \left(N_{b}-1\right)X$.
For a $\psi$-unitary QC-LDGM code, instead, the weight of the last $N-K$ columns of $\mathbf{G}$ is greater.
However, it remains significantly smaller than that of a generic code, whose
generator matrix is dense. In the latter case, (\ref{eq:Cenc}) gives an encoding complexity
approximately equal to $K/2$, and this number can be extremely large for rather long codes.

As for decoding complexity, we consider that belief propagation decoding algorithms work on
the Tanner graph, exchanging messages along its edges.
Thus, as a measure of decoding complexity, we can use the number of messages exchanged by belief
propagation per decoded bit per iteration, that coincides with the average Hamming weight 
of the parity-check matrix columns. In formula:

\begin{equation}
C_{dec} = \frac{\sum_{i=0}^{N-1} W\left[\mathbf{h}_i \right]}{N},
\label{eq:Cdec}
\end{equation}
where $W\left[\mathbf{h}_i \right]$ denotes the Hamming weight of the $i$-th column of $\mathbf{H}$.
In the parity-check matrix of a regular $\textit{I}$-based QC-LDGM code, the first $\left(N_b - 1\right)$ 
blocks have column weight $X$, while the last block has column weight $1$. So, for these codes,
$C_{dec} = {\left[\left(N_b - 1\right)X + 1\right]}/{N_b}$.

\subsection{Code examples}
\label{subsec:Examples}

In this section, we consider some examples of codes having parity-check matrices in the form (\ref{eq:HCircRow}) 
and compare the performance achieved by $\textit{I}$-based ($C_{\textit{I}}$) and $\psi$-unitary ($C_\psi$) QC-LDGM codes.

The parameters of the considered codes are summarized in Table \ref{tab:CodePars}.
Matrix $\mathbf{H}_{N_{b}-1}$ is specified by the positions of the non-zero coefficients in
its representative polynomial. The table also provides the weights of the other blocks of $\mathbf{H}$.
As expected, $\textit{I}$-based codes exhibit the lowest encoding complexity, but this is paid in terms of
error correction performance, as will be shown in the following.
This can also be argued by the minimum distance values, reported in Table \ref{tab:CodePars}, together with
the corresponding multiplicity. For $\textit{I}$-based codes, $d_{\min}$ simply coincides
with the lowest column weight in the leftmost part of the parity-check matrix ($K$ columns), augmented by $1$.
For $\psi$-unitary codes, instead, the minimum distance can be higher, and this reflects into lower error floors.
The minimum distance of the codes and their minimum weight codewords multiplicity have been estimated 
as explained in Section \ref{subsec:MinDist}.


\begin{table*}[t]
\caption{Parameters of the considered $\textit{I}$-based ($C_I$) and $\psi$-unitary ($C_\psi$) QC-LDGM codes}
\scriptsize
\begin{centering}
\begin{tabular}{c|l|l|l|l|l|l|l|l|l}
\hline
Case & Code($N$, $K$) 				& $n$ 		& $W[\mathbf{H}_i], i = 0, 1, \ldots, N_{b}-1$ & $\mathbf{H}_{N_{b}-1}$ & $W[\mathbf{H}_{N_{b}-1}^{-1}]$ & $d_{\min}$ & $A_{d_{\min}}$ & $C_{enc}$ 	& $C_{dec}$ \\
\hline
& $C_I(2560, 2048)$														& $512$	& $\left\{6,6,6,6,1\right\}$ & $(0)$ & $1$ & $7$ 	& $2048$	& $24$ & $5$ \\[-1ex]
\raisebox{1.5ex}{$1$} & $C_\psi(2560, 2048)$ 	& $512$ & $\left\{5,5,5,5,5\right\}$ & $(0; 8; 24; 72; 152)$ & $19$ & $10$ & $5120$ & $338$ & $5$ \\
\hline
& $C_I^a(1248, 936)$ 												& $312$ & $\left\{5,5,5,1\right\}$ & $(0)$ & $1$ & $6$ 	& $936$ & $15$ & $4$ \\
& $C_I^b(1248, 936)$ 												& $312$ & $\left\{7,6,6,1\right\}$ & $(0)$ & $1$ & $7$ 	& $624$ 	& $19$ & $5$ \\[-2ex]
\raisebox{2.5ex}{$2$} & $C_\psi(1248, 936)$ & $312$ & $\left\{5,5,5,5\right\}$ & $(0; 3; 9; 42; 87)$ & $21$ & $10$ & $1872$ & $261$ & $5$ \\
\hline
& $C_I^a(1880, 1504)$  											 & $376$ & $\left\{5,5,5,5,1\right\}$ & $(0)$ & $1$ & $6$ & $1504$ & $20$ & $4.2$ \\
& $C_I^b(1880, 1504)$  											 & $376$ & $\left\{6,6,6,6,1\right\}$ & $(0)$ & $1$ & $7$ & $1504$ & $24$ & $5$ \\[-2ex]
\raisebox{2.5ex}{$3$} & $C_\psi(1880, 1504)$ & $376$ & $\left\{5,5,5,5,5\right\}$ & $(0; 6; 18; 53; 112)$ & $21$ & $10$ & $3760$ & $366$ & $5$ \\
\hline
& $C_I^a(8192, 7168)$  											 & $1024$ & $\left\{5,5,5,5,5,5,5,1\right\}$ & $(0)$ & $1$ & $6$ & $7168$ & $35$ & $4.5$ \\
& $C_I^b(8192, 7168)$  											 & $1024$ & $\left\{6,6,6,6,5,5,5,1\right\}$ & $(0)$ & $1$ & $6$ & $3072$ & $39$ & $5$ \\[-2ex]
\raisebox{2.5ex}{$4$} & $C_\psi(8192, 7168)$ & $1024$ & $\left\{5,5,5,5,5,5,5,5\right\}$ & $(0; 32; 160; 224; 480)$ & $15$ & $10$ & $28672$ & $525$ & $5$ \\
\hline
& $C_I(10000, 5000)$													& $5000$ 	& $\left\{5,1\right\}$ 									& $(0)$ & $1$ & $6$ & $5000$ & $5$ & $3$ \\
& $C_I(5000, 4500)$ 													& $500$ 	& $\left\{4,4,3,3,3,3,3,3,3,1\right\}$ 	& $(0)$ & $1$ & $4$ & $3500$ & $29$ & $3$ \\ [-2ex]
\raisebox{2.5ex}{$5$} & $C_\psi(5000, 4500)$ 	& $500$ 	& $\left\{3,3,3,3,3,3,3,3,3,3\right\}$ 	& $(0; 3; 128)$ & $9$ & $6$ & $22500$ & $243$ & $3$ \\
\hline
\end{tabular}
\par\end{centering}
\label{tab:CodePars} 
\end{table*}

\subsubsection{Case $1$}
\label{subsubsec:Case1}

\begin{figure}
\begin{centering}
\subfigure[]{\includegraphics[width=43mm]{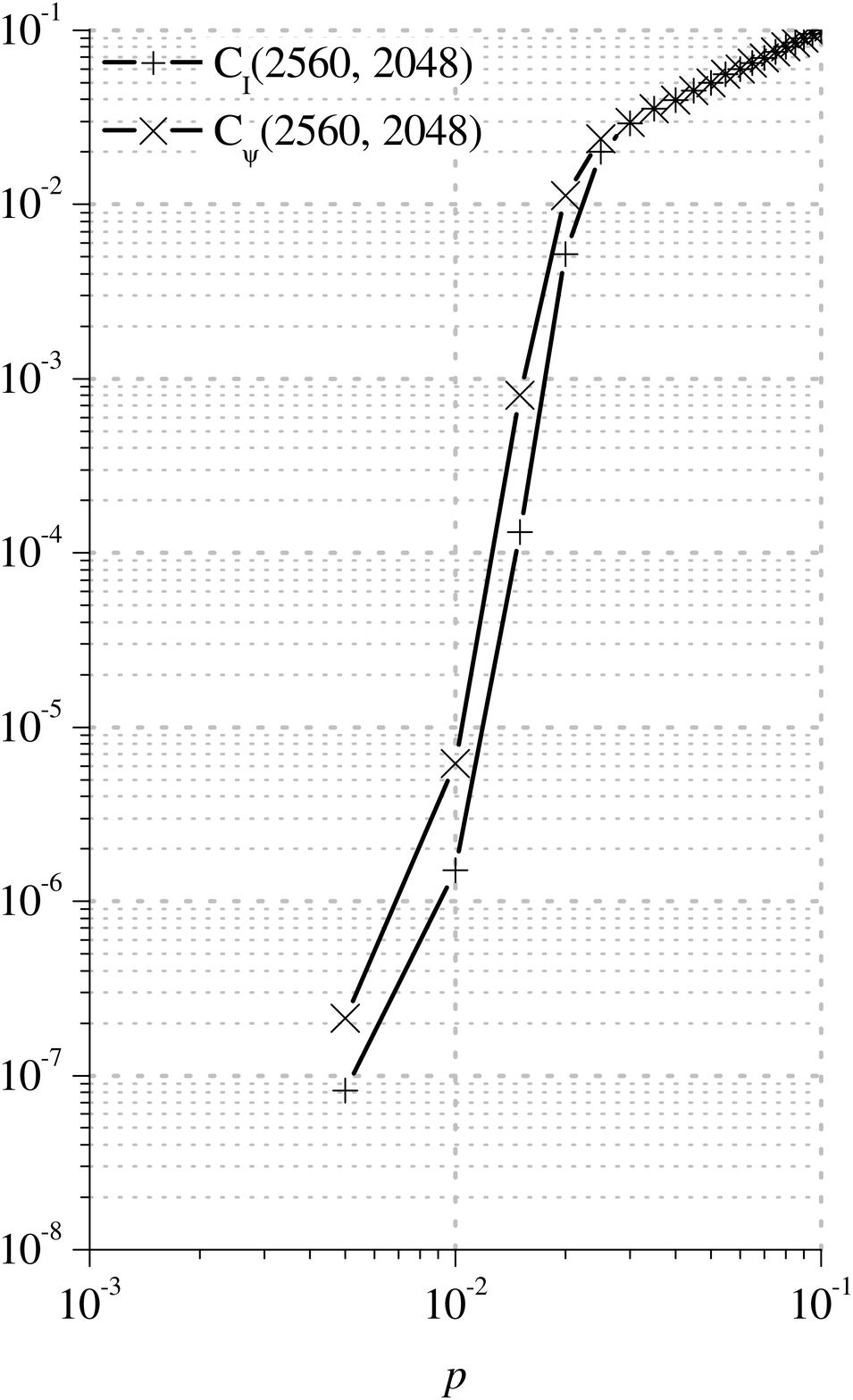}}
\subfigure[]{\includegraphics[width=43mm]{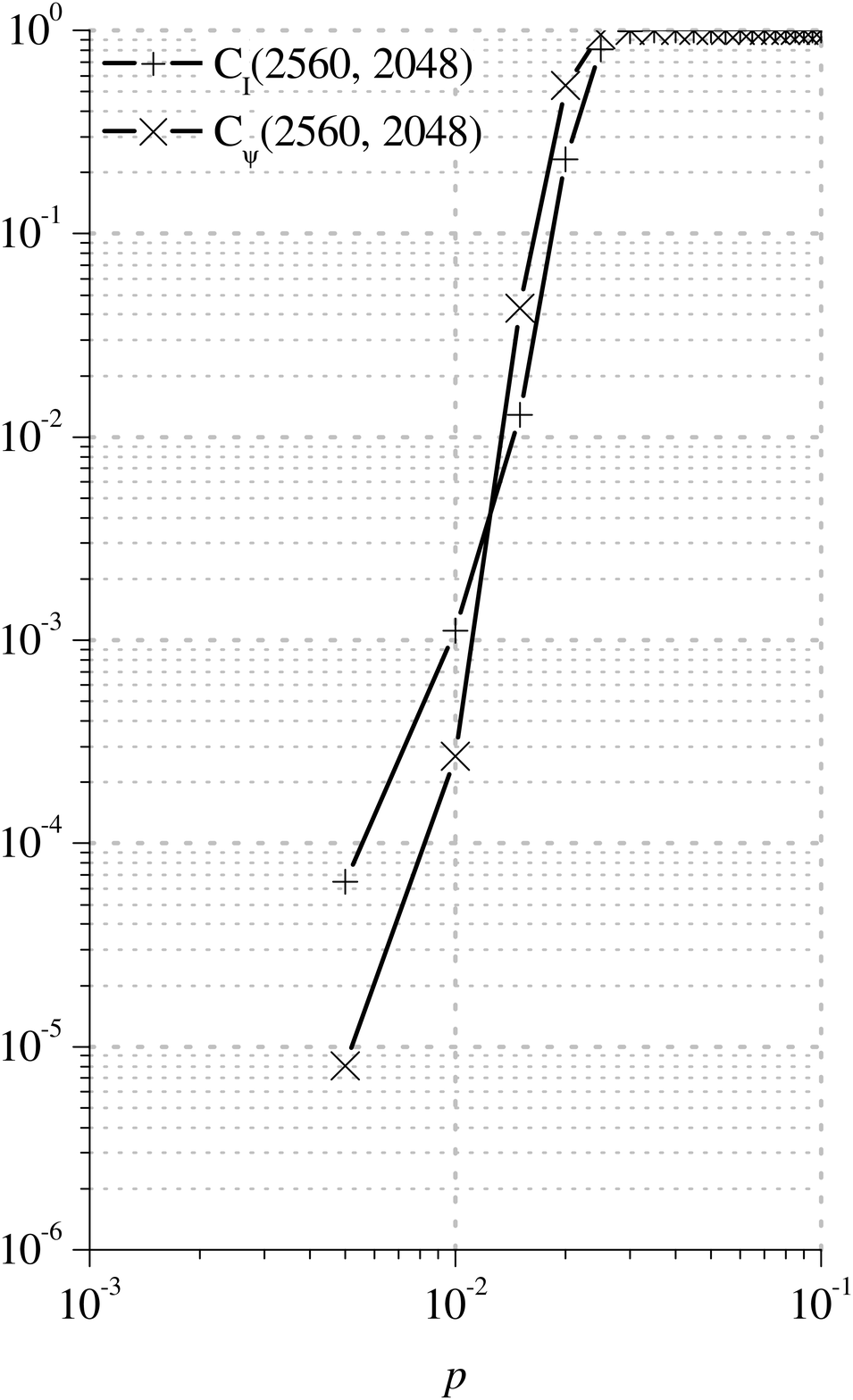}}
\par\end{centering}
\caption{(a) BER and (b) FER performance of QC-LDGM codes with $N = 2560$ and $K = 2048$ over the BSC.
\label{cap:FirstSim}}
\end{figure}

As a first example, we have considered two codes with length $N = 2560$ and dimension $K = 2048$, having
parity-check matrices formed by a row of $N_b = 5$ circulant blocks with size $512$.
Their error correction performance has been assessed over the BSC and Fig. \ref{cap:FirstSim} reports the
simulated BER and FER curves. 

The simulated FER curves confirm that the performance of the $\psi$-unitary code is
better with respect to that of the $\textit{I}$-based code, even for transition
probabilities within the simulation scope.
The same is not equally evident in the BER curves, since their intersection should occur
for very low error rates.

The improvement in performance is paid in terms of complexity. In fact, though both codes have the same decoding
complexity, the $\psi$-unitary code exhibits higher encoding complexity.
Thanks to its LDGM character, however, encoding complexity is still considerably smaller 
than that of a generic code, that, in this case, would be, approximately, $K/2 = 1024$.

\subsubsection{Case $2$}
\label{subsubsec:Case2}

\begin{figure}
\begin{centering}
\subfigure[]{\includegraphics[width=43mm]{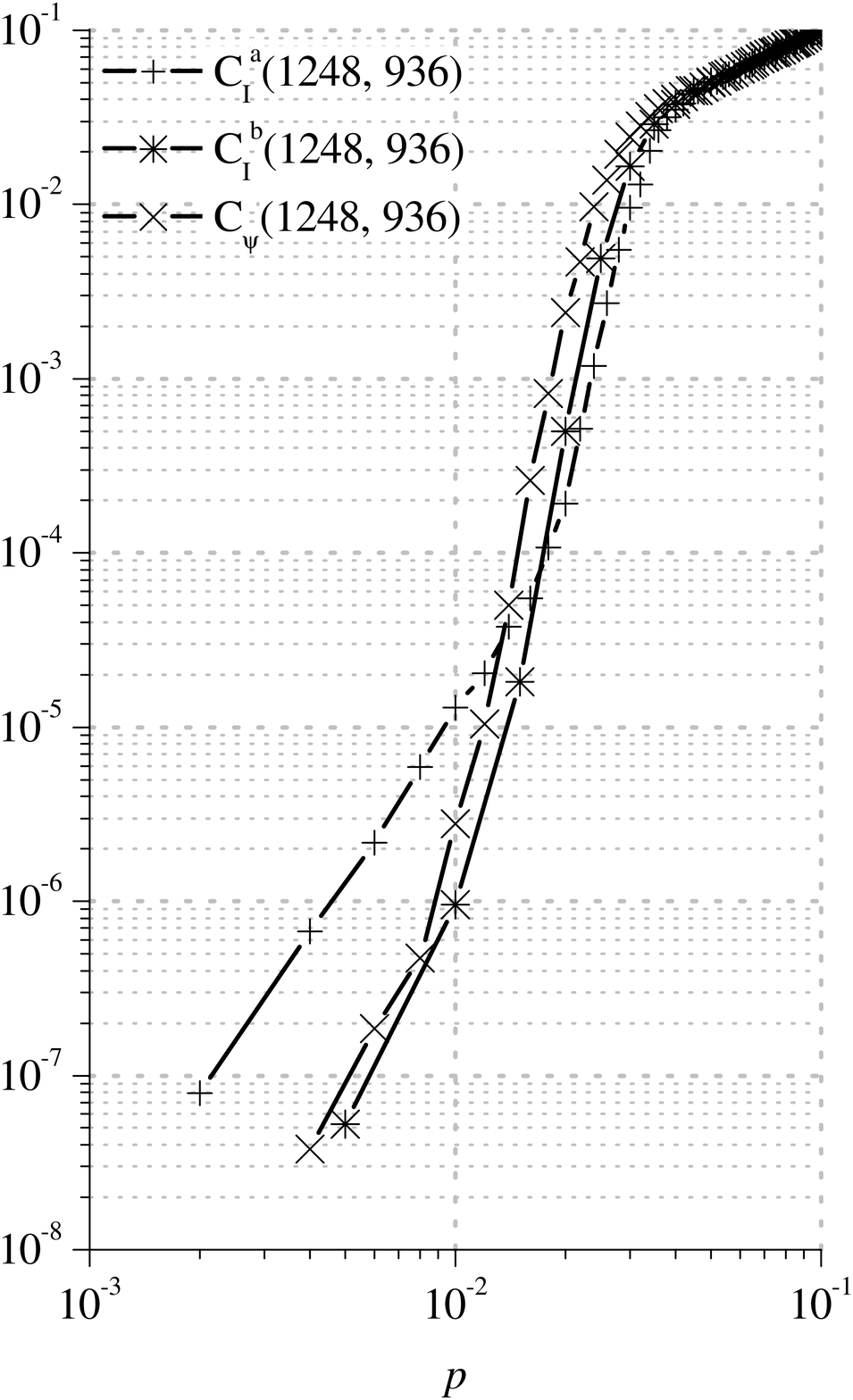}}
\subfigure[]{\includegraphics[width=43mm]{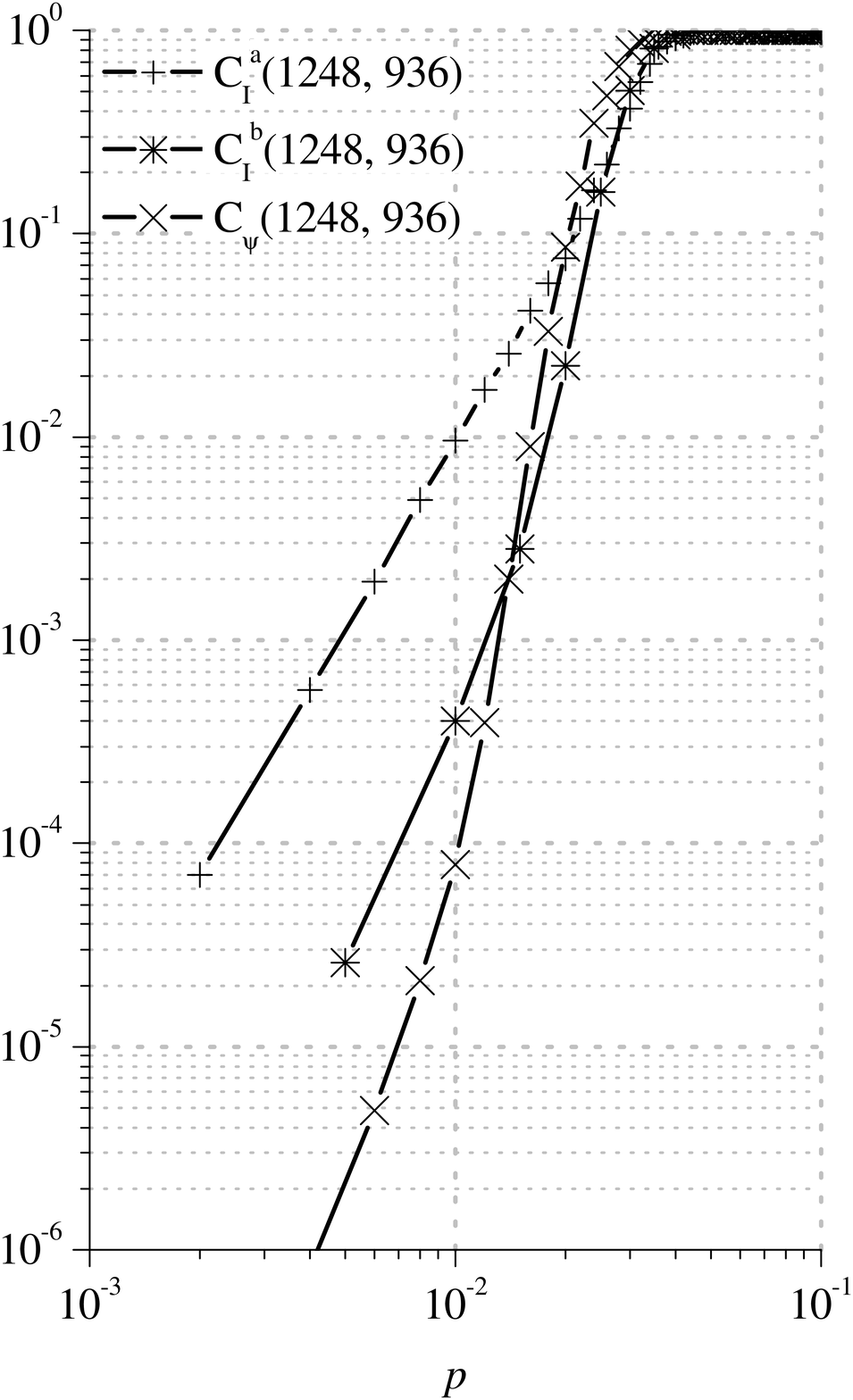}}
\par\end{centering}
\caption{(a) BER and (b) FER performance of QC-LDGM codes with $N = 1248$ and $K = 936$ over the BSC.
\label{cap:SimBSC}}
\end{figure}

As a second example, we have considered the parameters of the rate $3/4$ "A" IEEE 802.16e standard LDPC code with $z$ factor $52$ \cite{802.16e}, that has length $N = 1248$ and dimension $K = 936$.
A QC-LDGM code has been designed, denoted as $C_I^a(1248, 936)$, that has a parity-check matrix formed by three $312 \times 312$ circulant blocks with weight $X=5$, followed by a $312 \times 312$ identity matrix. We have then considered a second $\textit{I}$-based code, denoted as $C_I^b(1248, 936)$, that has $3$ blocks with higher weight, in such a way as to increase its minimum distance.
In order to compare the error rate performance and complexity of these two $\textit{I}$-based QC-LDGM codes with those of a $\psi$-unitary code, we have designed the code denoted as $C_{\psi}(1248, 936)$, having a parity-check matrix formed by all weight-$5$ blocks, the last of which is a $312 \times 312$ $\psi$-unitary circulant matrix.
Fig. \ref{cap:SimBSC} reports the simulated performance of these codes over the BSC.
As for other examples in the following, the first $\textit{I}$-based QC-LDGM code has the same weight of non-identity blocks as the $\psi$-unitary code. The second $\textit{I}$-based QC-LDGM code, instead, has non-identity blocks with increased weight in order to have the same decoding complexity as the $\psi$-unitary code.

\begin{figure}
\begin{centering}
\subfigure[]{\includegraphics[width=43mm]{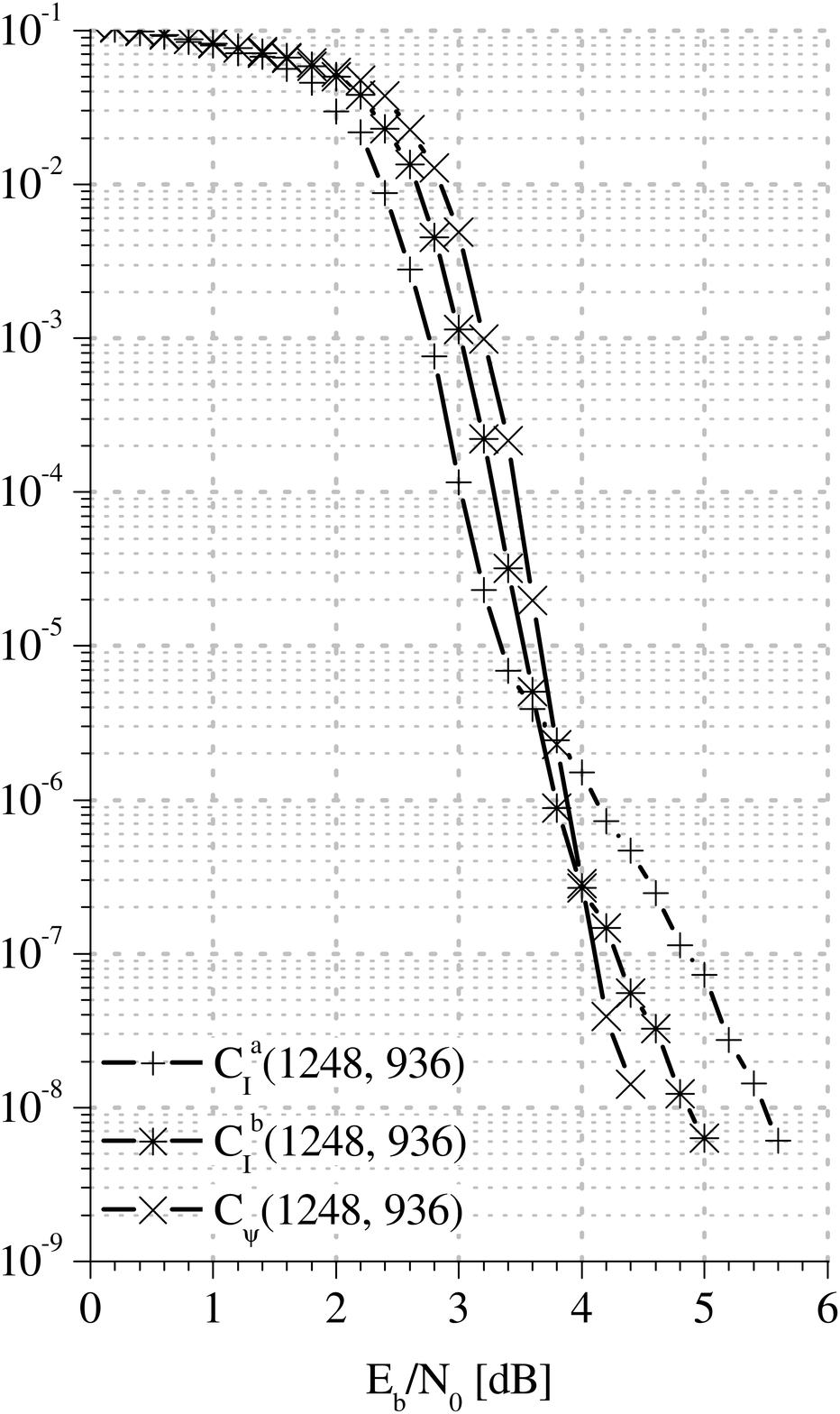}}
\subfigure[]{\includegraphics[width=43mm]{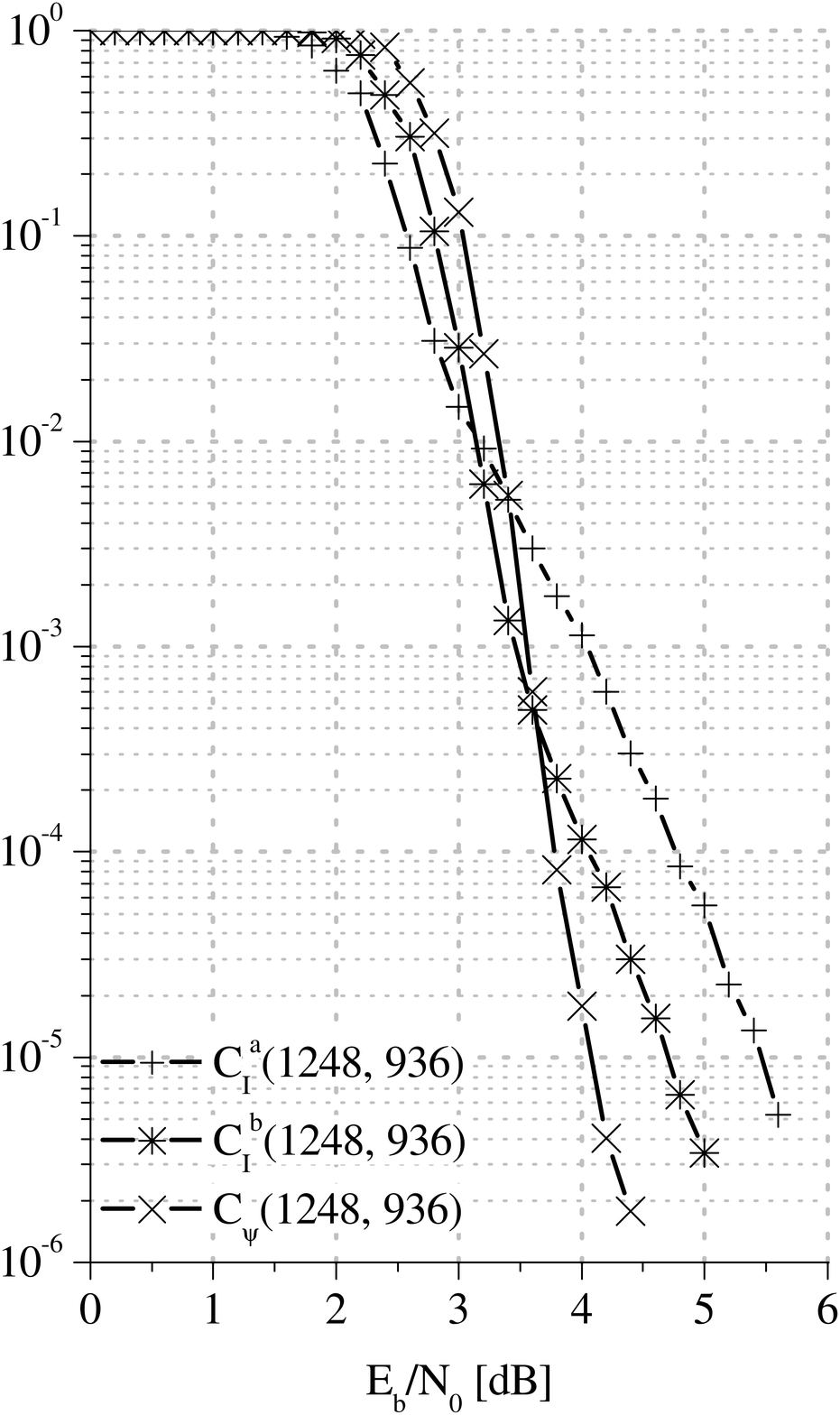}}
\par\end{centering}
\caption{(a) BER and (b) FER performance of QC-LDGM codes with $N = 1248$ and $K = 936$ over the AWGN channel.
\label{cap:Sim80216e}}
\end{figure}

From Fig. \ref{cap:SimBSC} we see that the two $\textit{I}$-based QC-LDGM codes have rather high error floors.
The code designed by using the $\psi$-unitary matrix, instead, is able to achieve better performance in the error floor region.
This results from the FER curve, showing a more favorable slope.
The same is not equally evident in the BER curve, that, however, exhibits a better slope when transmission is simulated over the AWGN channel (see Fig. \ref{cap:Sim80216e}).


\subsubsection{Case $3$}
\label{subsubsec:Case3}

Another relevant example can be obtained by considering the parameters of the codes used in the Digital Video Broadcasting - Return Channel Satellite (DVB-RCS) standard \cite{DVB-RCS}. Actually, the current version of the standard uses a turbo code and not an LDPC code. The turbo encoder consists of a double binary circular recursive systematic convolutional code, combined with an optimized two-level interleaver and a puncturing map to deal with variable rates. However, the possibility to replace the turbo code with LDPC codes has been explored in recent literature, and encouraging results have been obtained \cite{Lestable2006}, \cite{Baldi2009}.

\begin{figure}
\begin{centering}
\subfigure[]{\includegraphics[width=43mm]{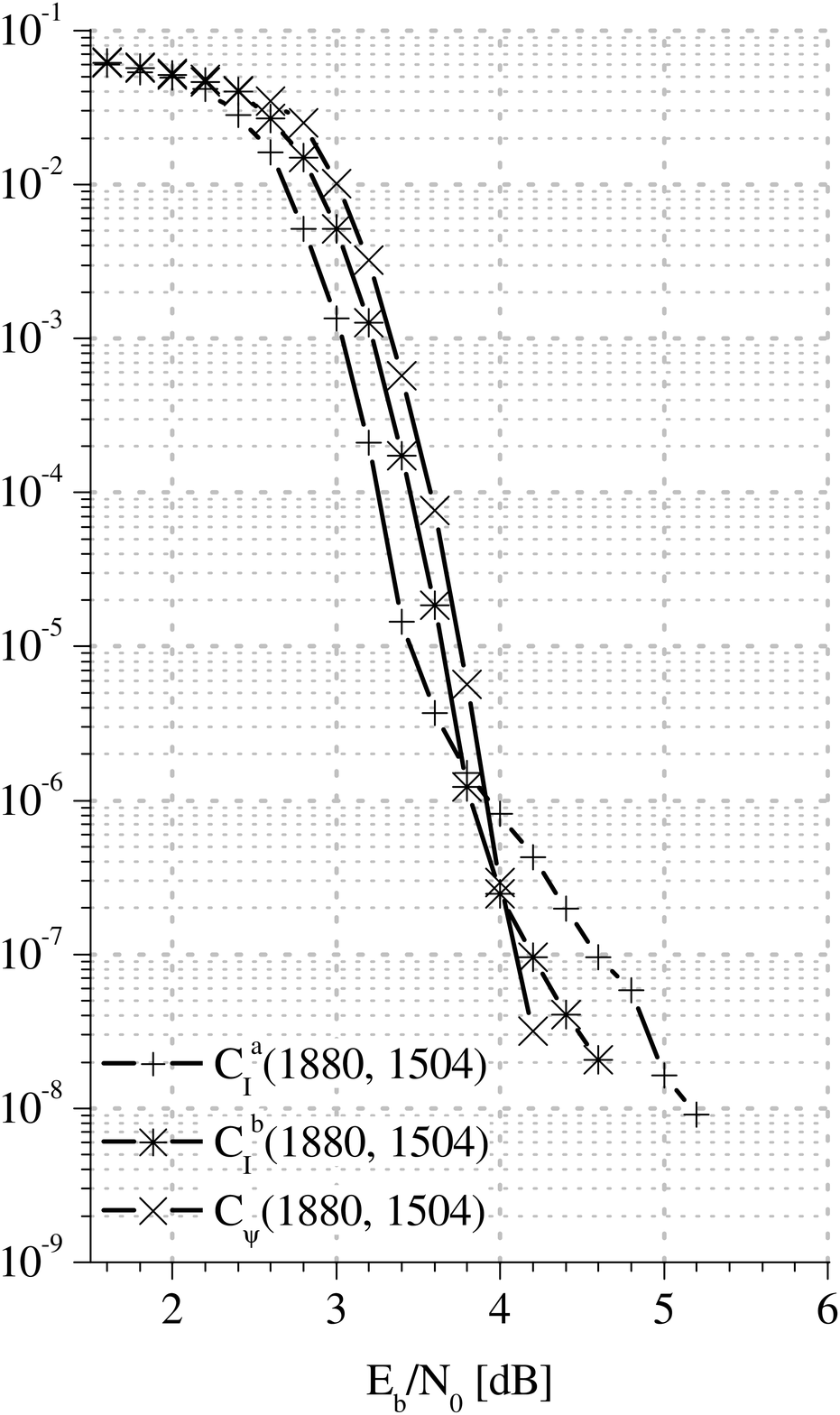}}
\subfigure[]{\includegraphics[width=43mm]{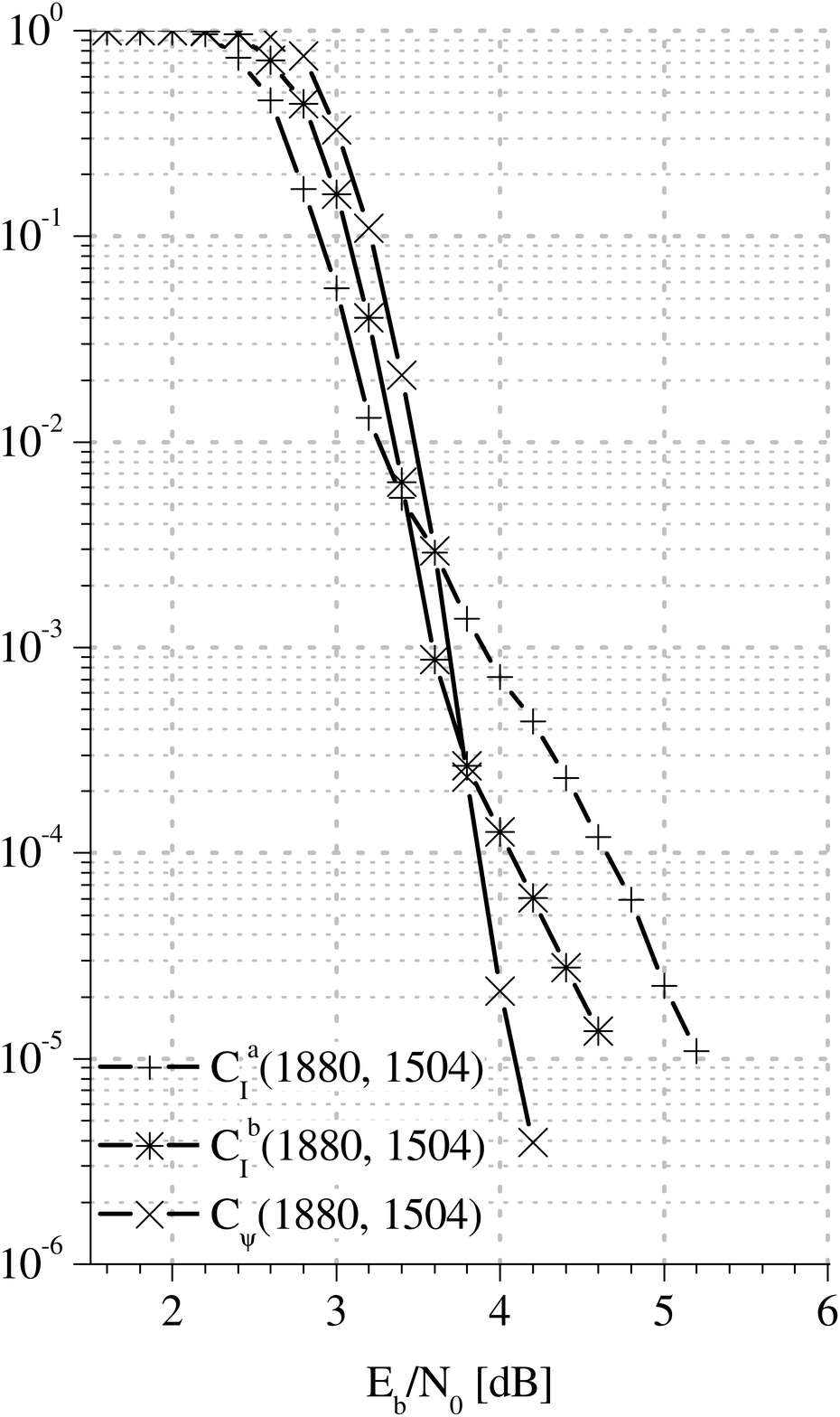}}
\par\end{centering}
\caption{(a) BER and (b) FER performance of QC-LDGM codes with $N = 1880$ and $K = 1504$ over the AWGN channel.
\label{cap:SimDVBRCS}}
\end{figure}

Figure \ref{cap:SimDVBRCS} shows the performance of three codes with MPEG2 information block size (that is 188 bytes) and code rate $4/5$. 
For all codes, the parity-check matrix has the structure (\ref{eq:HCircRow}), with $N_b = 5$ and $n = 376$; so the code has length $N = 1880$ and dimension $K = 1504$.
The first code is an $\textit{I}$-based QC-LDGM code and it is denoted as $C_I^a(1880, 1504)$. Its parity-check matrix is formed by four circulant blocks, $\mathbf{H}_0, \ldots, \mathbf{H}_3$, with weight $X=5$, followed by an identity matrix.
The second code, $C_I^b(1880, 1504)$, is still an $\textit{I}$-based QC-LDGM code, but its parity-check matrix includes weight-$6$ blocks.
The third code is instead a $\psi$-unitary QC-LDGM code, with a parity-check matrix formed by $N_b = 5$ circulant blocks with weight $5$, the last of which is a $\psi$-unitary circulant matrix.
From the figure we see that, also in this case, $\textit{I}$-based QC-LDGM codes exhibit a significant error floor, while the $\psi$-unitary code has simulated curves with a more favorable slope for increasing signal-to-noise ratio.

\subsubsection{Case $4$}
\label{subsubsec:Case4}

\begin{figure}
\begin{centering}
\subfigure[]{\includegraphics[width=43mm]{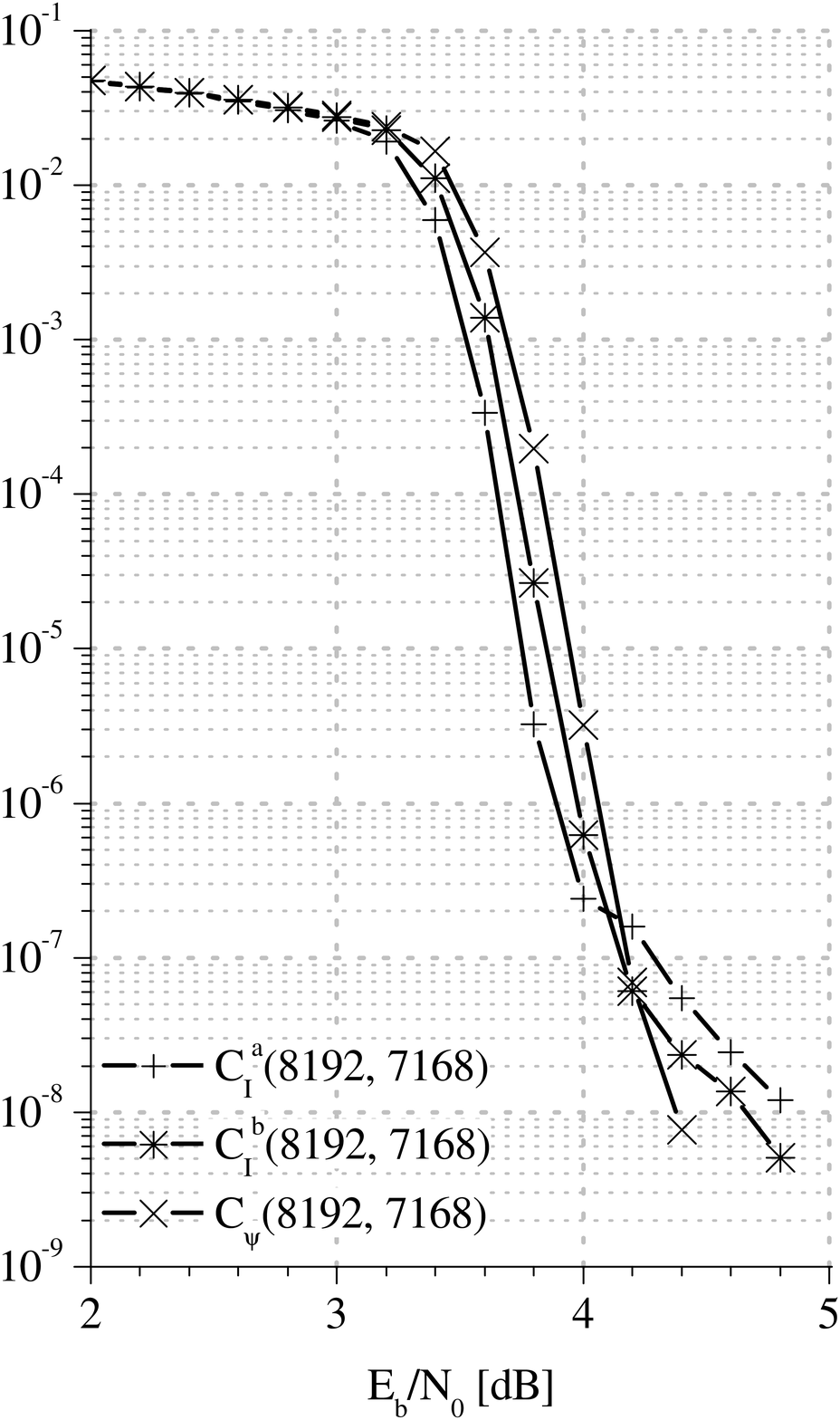}}
\subfigure[]{\includegraphics[width=43mm]{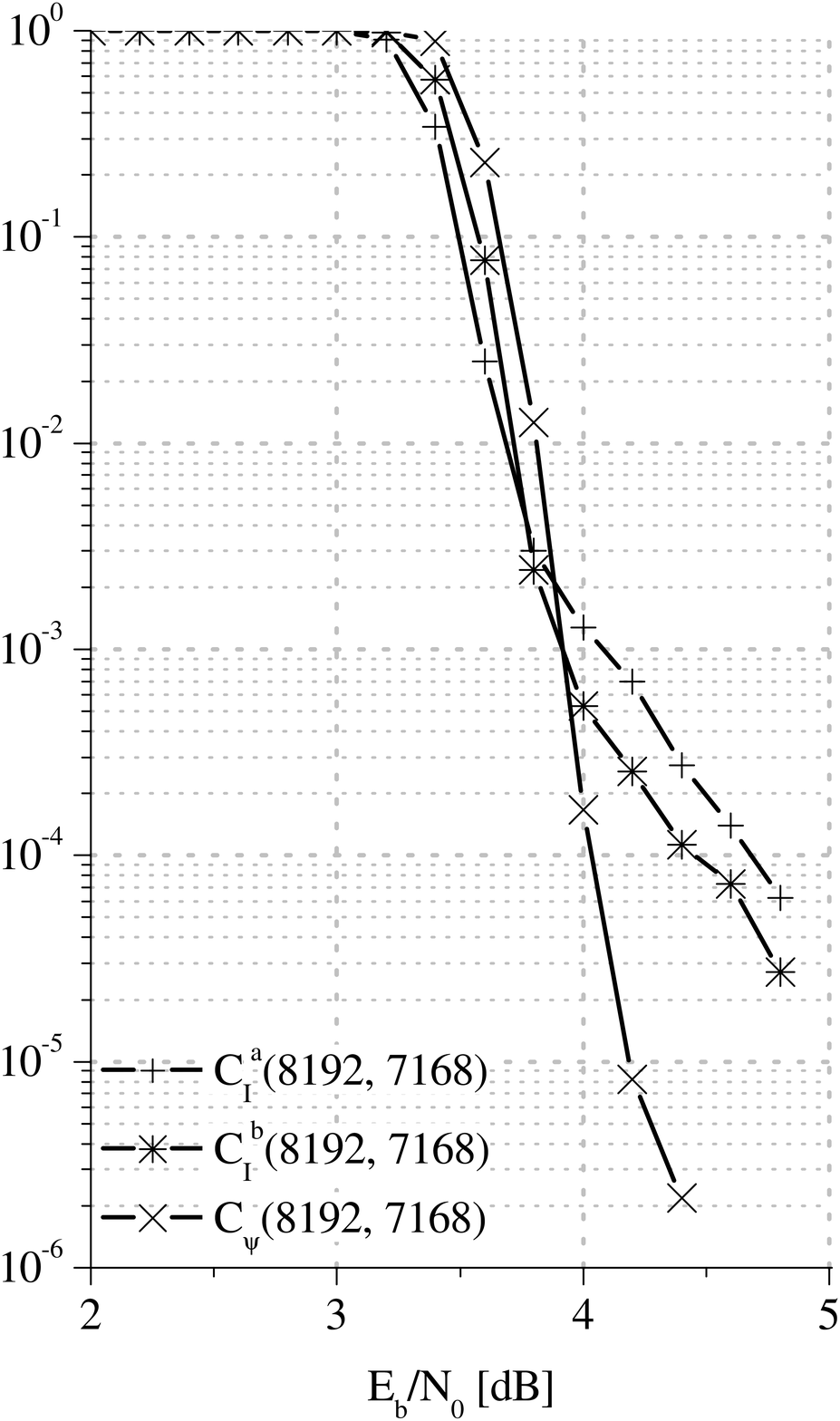}}
\par\end{centering}
\caption{(a) BER and (b) FER performance of QC-LDGM codes with $N = 8192$ and $K = 7168$ over the AWGN channel.
\label{cap:SimCCSDS}}
\end{figure}

A fourth example is shown in Fig. \ref{cap:SimCCSDS} for codes with $N = 8192$, $K = 7168$ and code rate $7/8$. These parameters are of interest, as they are very close to those adopted in a well known LDPC code proposed for near-Earth space missions by the Consultative Committee for Space Data Systems (CCSDS) \cite{CCSDS2007}. 
We compare the performance of three QC-LDGM codes having parity-check matrices formed by a row of $8$ circulant blocks with size $n=1024$. 
The first code, $C_I^a(8192, 7168)$, is an $\textit{I}$-based QC-LDGM code with all non-identity blocks having weight $5$. The second code, $C_I^b(8192, 7168)$, is also an $\textit{I}$-based QC-LDGM code, but its parity-check matrix includes weight-$6$ blocks.
The third code is instead a $\psi$-unitary QC-LDGM code, with a parity-check matrix formed by all weight-$5$ blocks.
From the figure we see that the latter code has good performance, with steep slope of the curves and low error floor. The two $\textit{I}$-based codes, on the contrary, exhibit a rather high error floor.


\subsubsection{Case $5$}
\label{subsubsec:Case5}

A common solution for reducing the error floor in LDGM codes is represented by the adoption of concatenated schemes, formed by an outer high-rate LDGM code followed by an inner low-rate LDGM code \cite{Garcia-Frias2003}, \cite{Gonzalez-Lopez2007}.
Decoding is accomplished by using two belief propagation decoders in serial concatenation: first, the inner code is decoded starting from channel information, then the \textit{a posteriori} messages it produces are used as \textit{a priori} information to initialize the outer decoder.

By this procedure, the error floor of a single low-rate LDGM code can be often significantly reduced, at the cost of increased complexity due to the serial concatenation.
However, we have verified that, also in concatenated schemes, the adoption of $\psi$-unitary codes can help to improve performance.
As an example, we have considered a $(10000, 5000)$ inner QC-LDGM code having parity-check matrix formed by a $5000 \times 5000$ circulant block with weight $5$, followed by a $5000 \times 5000$ identity matrix. Such code, denoted as $C_I(10000, 5000)$ in Table \ref{tab:CodePars}, has been used in serial concatenation with two different $(5000, 4500)$ outer codes, both described by a parity-check matrix in the form of a row of ten $500 \times 500$ circulant blocks. The first outer code, denoted as $C_I(5000, 4500)$, is an $\textit{I}$-based QC-LDGM code, characterized by: two circulant blocks with weight $4$, seven circulant blocks with weight $3$ and one identity block. The second code, denoted as $C_\psi(5000, 4500)$, is described by $10$ circulant blocks with weight $3$, the last one being a $\psi$-unitary block. Their parameters have been chosen in such a way as to obtain the same decoding complexity for both codes.

\begin{figure}
\begin{centering}
\subfigure[]{\includegraphics[width=43mm]{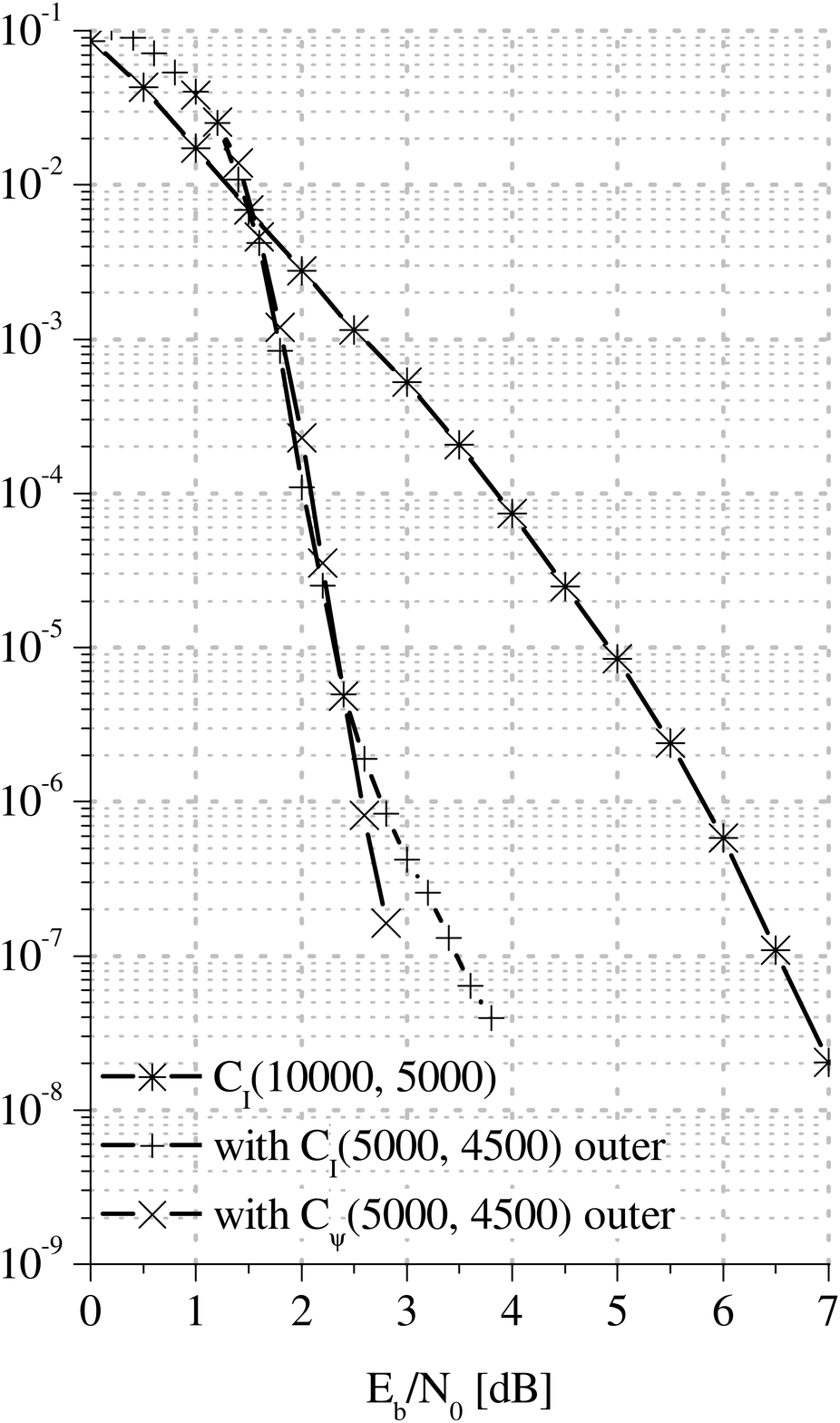}}
\subfigure[]{\includegraphics[width=43mm]{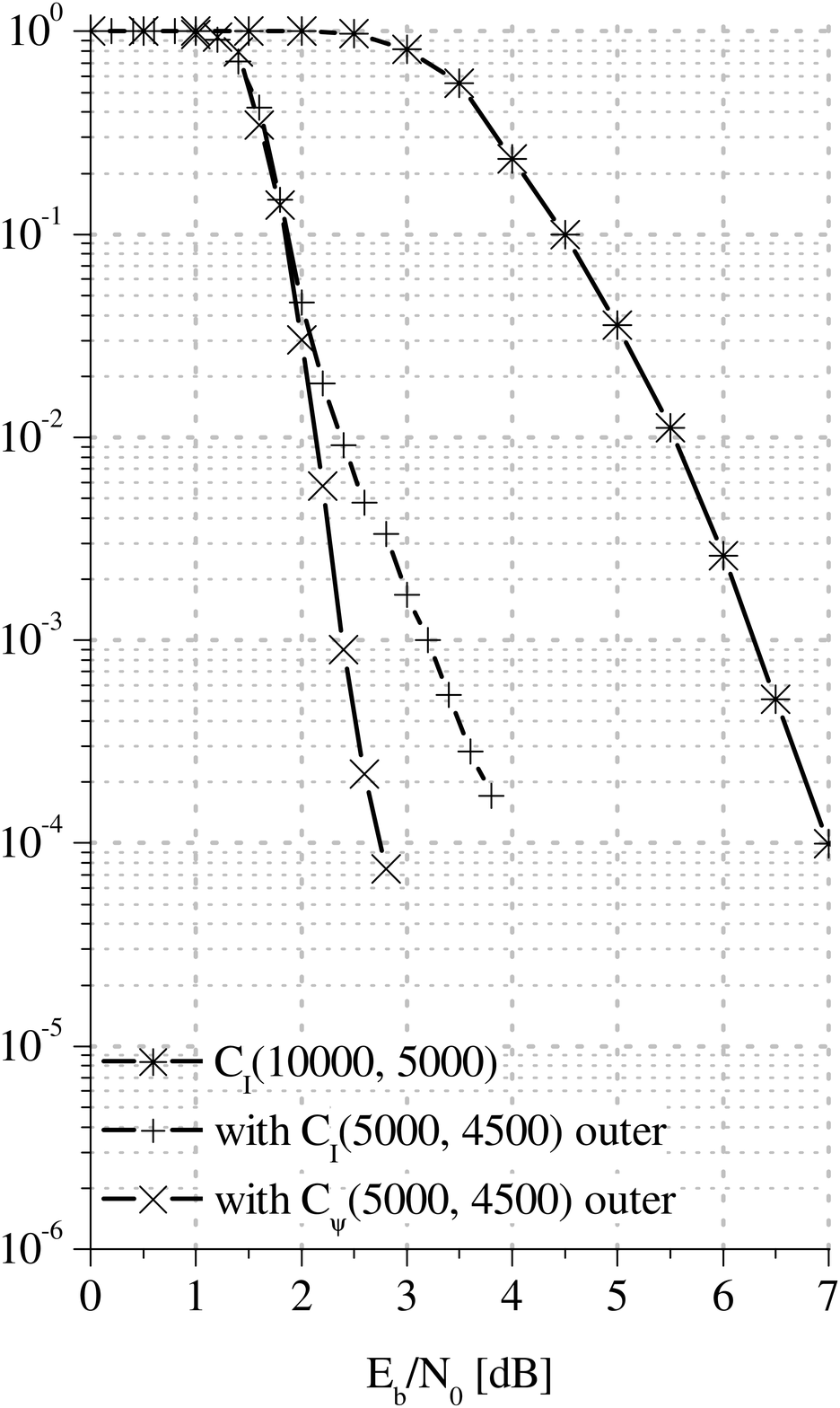}}
\par\end{centering}
\caption{(a) BER and (b) FER performance over the AWGN channel of QC-LDGM codes of Case $5$.
\label{cap:SimConc}}
\end{figure}

As we notice from Fig. \ref{cap:SimConc}, code $C_I(10000, 5000)$, when used alone, has quite poor error correction performance, and the adoption of the concatenated scheme actually allows to improve it.
On the other hand, simulations show that including the $\psi$-unitary code in the concatenated scheme has positive effects on the error floor: it permits to improve further the performance with respect to the adoption of an outer $\textit{I}$-based QC-LDGM code.

\section{Conclusion}
\label{sec:Cinque}

The first goal of this paper was to define a new class of sparse circulant matrices, named $\psi$-unitary matrices, that are easily invertible and whose inverse can be sparse, though not being permutation matrices.
We have shown that, under suitable choices, these matrices can be free of length-$4$ cycles.

These features make the $\psi$-unitary matrices a smart element for the design of QC-LDGM codes, that are able to join the inner structure of quasi-cyclic codes with the existence of a sparse representation of the generator matrix. This ensures low complexity but also error floor performance better than that offered by other codes of the same class.

We have shown that good codes based on $\psi$-unitary matrices can be designed, with the code length and rate adopted for relevant applications like WiMax, DVB-RCS and space missions.

\appendices


\section{A bound on the weight of the inverse for $m > 1$.}
\label{app:1}
\begin{The}
\label{The235}
For $n = 2^{m+2}s$, $m>1$, the matrices in $\Xi^{2^{m+2}s}_s$ with $W[a] = 2m + 3$, whose inverse can be computed by using Theorem \ref{The232}, have
\begin{equation}
W[a^{-1}] \leq 11 (2m + 3)\prod_{k=2}^m(2k+3).
\label{UpInvBound}
\end{equation}

\begin{IEEEproof}
Let us consider (\ref{eq:Inverter}). Based on the expressions in Section \ref{F_inv}, we know that:
\[W[a^{2^m} + w] \leq 11.\]
On the other hand, it is easy to find:
\begin{equation}
  W[a^{2^i}] = \left \{
    \begin{array}{ll}
    2m + 3 & {\rm for}\; i = 0,\\
    2m+5-2i & {\rm for}\; i > 0.
    \end{array} \right.
\end{equation}
Therefore:
\begin{eqnarray*}
W[a^{-1}] & \leq & W[a^{2^m} + w]\prod_{i = 0}^{m - 1} W[a^{2^{i}}] \\
& \leq & 11 (2m+3) \prod_{i=1}^{m-1}(2m+5-2i)
\end{eqnarray*}
and, through simple algebra, (\ref{UpInvBound}) is finally obtained. 

\end{IEEEproof}
\end{The}

It should be noticed that the upper bound (\ref{UpInvBound}) can be loose: depending on the matrix structure, the actual weight of the inverse can be much smaller.

\begin{example}
Let us consider the matrix of the Example \ref{ex:p4}, that is characterized by $m = 2$. Its inverse, computed through (\ref{eq:Inverter}), results in:
\begin{eqnarray*}
a^{-1}(x) & = & (a^4(x) + w(x))a^2(x) a(x) \\
					&	=	& (0;1;2;8;12;14;15;16;17;19;20;24;28;32;36; \\
					&		& 39;40;43;44;48;56;60;63;68;72;80;84;87;92; \\
					&		& 96;103;105;107;108;120;127;131;132;144;151; \\
					&		& 156;168;175)_{176}
\end{eqnarray*}
and, therefore, $W[a^{-1}] = 43$. On the other hand, the upper bound in this case gives $W[a^{-1}]<539$, that exceeds the value of $n$. It should be observed, however, that the polynomial $a(x)$ (as well as $a^{-1}(x)$) can be scaled in such a way as to maintain the weight with a larger size. For example, we can define:
\begin{eqnarray*}
a'(x) & = & 8 \ast a(x) = 8 \ast (0; 1; 3; 7; 12; 25; 51)_{176} = \\ 
			&	=	& (0; 8; 24; 56; 96; 200; 408)_{1408}
\end{eqnarray*}
such that $a'(x)$ has the same features of $a(x)$ but a lower density. Moreover, its inverse has the same weight of $a^{-1}(x)$.
\end{example}

It is also meaningful to compare the upper bound (\ref{UpInvBound}) with the true upper bound determined through a numerical search. Examples are shown in Table \ref{tab:p51b} for $2 \leq m \leq 4$.

\begin{center}
\begin{table}[ht]
\caption{Estimated and true upper bound on the weight of the inverse matrix for some values of $m$}
\begin{centering}
\begin{tabular}{|c|cc|}
	\hline
$m$ & Upper bound estimate (\ref{UpInvBound}) & True upper bound \\
	\hline
2 & 539 & 269 \\
3 & 6237 & 1873  \\
4 & 83853 & 14969 \\
	\hline
\end{tabular}
\par\end{centering}
\label{tab:p51b} 
\end{table}
\par\end{center}

As already observed for $m = 1$, the upper bound becomes smaller if specific relationships between the $k_i$'s are established. As an example, for $m = 2, k_1 = 3k_0$ and $k_2 = 4k_0$ we find $W[a^{-1}] \leq 75$.

\section*{Acknowledgment}
The authors are grateful to the anonymous reviewers for their comments,
which helped to improve several aspects of this paper.
In particular, they wish to thank one reviewer for having suggested the procedure 
for obtaining the minimum distance and multiplicity of the considered codes.
They are also very grateful to Prof. Igal Sason for fruitful discussion on the
estimate of the codes performance.

\bibliographystyle{ieeetran}

\begin{thebibliography}{10}
\providecommand{\url}[1]{#1}
\csname url@samestyle\endcsname
\providecommand{\newblock}{\relax}
\providecommand{\bibinfo}[2]{#2}
\providecommand{\BIBentrySTDinterwordspacing}{\spaceskip=0pt\relax}
\providecommand{\BIBentryALTinterwordstretchfactor}{4}
\providecommand{\BIBentryALTinterwordspacing}{\spaceskip=\fontdimen2\font plus
\BIBentryALTinterwordstretchfactor\fontdimen3\font minus
  \fontdimen4\font\relax}
\providecommand{\BIBforeignlanguage}[2]{{%
\expandafter\ifx\csname l@#1\endcsname\relax
\typeout{** WARNING: IEEEtran.bst: No hyphenation pattern has been}%
\typeout{** loaded for the language `#1'. Using the pattern for}%
\typeout{** the default language instead.}%
\else
\language=\csname l@#1\endcsname
\fi
#2}}
\providecommand{\BIBdecl}{\relax}
\BIBdecl

\bibitem{Richardson2001}
T.~Richardson and R.~Urbanke, ``The capacity of low-density parity-check codes
  under message-passing decoding,'' \emph{{IEEE} Trans. Inform. Theory},
  vol.~47, no.~2, pp. 599--618, Feb. 2001.

\bibitem{Wiechman2007}
G.~Wiechman and I.~Sason, ``Parity-check density versus performance of binary
  linear block codes: New bounds and applications,'' \emph{{IEEE} Trans.
  Inform. Theory}, vol.~53, no.~2, pp. 550--579, Feb. 2007.

\bibitem{Richardson2001EfficientEncoding}
T.~Richardson and R.~Urbanke, ``Efficient encoding of low-density parity-check
  codes,'' \emph{{IEEE} Trans. Inform. Theory}, vol.~47, no.~2, pp. 638--656,
  Feb. 2001.

\bibitem{Freundlich2007}
S.~Freundlich, D.~Burshtein, and S.~Litsyn, ``Approximately lower triangular
  ensembles of {LDPC} codes with linear encoding complexity,'' \emph{{IEEE}
  Trans. Inform. Theory}, vol.~53, no.~4, pp. 1484--1494, Apr. 2007.

\bibitem{Haley2002}
D.~Haley, A.~Grant, and J.~Buetefuer, ``Iterative encoding of low-density
  parity-check codes,'' in \emph{Proc. {IEEE} Global Telecommunications
  Conference {(GLOBECOM '02)}}, vol.~2, Taipei, Taiwan, Nov. 2002, pp.
  1289--1293.

\bibitem{Haley2005}
D.~Haley and A.~Grant, ``Improved reversible {LDPC} codes,'' in \emph{Proc.
  {IEEE} International Symposium on Information Theory {(ISIT 2005)}},
  Adelaide, Australia, Sep. 2005, pp. 1367--1371.

\bibitem{Cheng1996}
J.~F. Cheng and R.~J. McEliece, ``Some high-rate near capacity codecs for the
  {Gaussian} channel,'' in \emph{Proc. 34th Allerton Conference on
  Communications, Control and Computing}, Allerton, IL, Oct. 1996.

\bibitem{Garcia-Frias2003}
J.~Garcia-Frias and W.~Zhong, ``Approaching {Shannon} performance by iterative
  decoding of linear codes with low-density generator matrix,'' \emph{{IEEE}
  Commun. Lett.}, vol.~7, no.~6, pp. 266--268, Jun. 2003.

\bibitem{Gonzalez-Lopez2007}
M.~González-López, F.~J. Vázquez-Araújo, L.~Castedo, and J.~Garcia-Frias,
  ``Serially-concatenated low-density generator matrix ({SCLDGM}) codes for
  transmission over {AWGN} and {Rayleigh} fading channels,'' \emph{{IEEE}
  Trans. Wireless Commun.}, vol.~6, no.~8, pp. 2753--2758, Aug. 2007.

\bibitem{Kim2006}
J.-S. Kim and H.-Y. Song, ``Concatenated {LDGM} codes with single decoder,''
  \emph{{IEEE} Commun. Lett.}, vol.~10, no.~4, pp. 287--289, Apr. 2006.

\bibitem{Oenning2001}
T.~R. Oenning and J.~Moon, ``A low-density generator matrix interpretation of
  parallel concatenated single bit parity codes,'' \emph{{IEEE} Trans. Magn.},
  vol.~37, no.~2, pp. 737--741, Mar. 2001.

\bibitem{Hsu2005}
C.-H. Hsu and A.~Anastasopoulos, ``Asymptotic weight distributions of irregular
  repeat-accumulate codes,'' in \emph{Proc. {IEEE} Global Telecommunications
  Conference {(GLOBECOM '05)}}, Saint Louis, MO, Nov. 2005, pp. 1147--1151.

\bibitem{Jin2000}
H.~Jin, A.~Khandekar, and R.~McEliece, ``Irregular repeat-accumulate codes,''
  in \emph{Proc. Second International Symposium on Turbo Codes}, Brest, France,
  Sep. 2000, pp. 1--8.

\bibitem{Abbasfar2004}
A.~Abbasfar, D.~Divsalar, and K.~Yao, ``Accumulate repeat accumulate codes,''
  in \emph{Proc. {IEEE} Global Telecommunications Conference {(GLOBECOM '04)}},
  Dallas, TX, Nov. 2004, pp. 509--513.

\bibitem{Pfister2007}
H.~D. Pfister and I.~Sason, ``Accumulate-repeat-accumulate codes:
  Capacity-achieving ensembles of systematic codes for the erasure channel with
  bounded complexity,'' \emph{{IEEE} Trans. Inform. Theory}, vol.~53, no.~2,
  pp. 2088--2115, Jun. 2007.

\bibitem{Li2006}
Z.~Li, L.~Chen, L.~Zeng, S.~Lin, and W.~Fong, ``Efficient encoding of
  quasi-cyclic low-density parity-check codes,'' \emph{{IEEE} Trans. Commun.},
  vol.~54, no.~1, pp. 71--81, Jan. 2006.

\bibitem{Fossorier2004}
M.~P.~C. Fossorier, ``{Q}uasi-cyclic low-density parity-check codes from
  circulant permutation matrices,'' \emph{{IEEE} Trans. Inform. Theory},
  vol.~50, no.~8, pp. 1788--1793, Aug. 2004.

\bibitem{802.16e}
802.16e 2005, \emph{{IEEE} {S}tandard for {L}ocal and {M}etropolitan {A}rea
  {N}etworks - {P}art 16: {A}ir {I}nterface for {F}ixed and {M}obile
  {B}roadband {W}ireless {A}ccess {S}ystems - {A}mendment for {P}hysical and
  {M}edium {A}ccess {C}ontrol {L}ayers for {C}ombined {F}ixed and {M}obile
  {O}peration in {L}icensed {B}ands}, {IEEE} Std., Dec. 2005.

\bibitem{Yoon2005}
C.~Yoon, E.~Choi, M.~Cheong, and S.-K. Lee, ``Arbitrary bit generation and
  correction technique for encoding {QC-LDPC} codes with dual-diagonal parity
  structure,'' in \emph{Proc. {IEEE} {WCNC} 2007}, Hong Kong, Mar. 2007, pp.
  663--667.

\bibitem{Myung2005}
S.~Myung, K.~Yang, and J.~Kim, ``Quasi-cyclic {LDPC} codes for fast encoding,''
  \emph{{IEEE} Trans. Inform. Theory}, vol.~51, no.~8, pp. 2894--2901, Aug.
  2005.

\bibitem{CCSDS2007}
{CCSDS}, ``{L}ow {D}ensity {P}arity {C}heck {C}odes for {U}se in {N}ear-{E}arth
  and {D}eep {S}pace {A}pplications,'' {C}onsultative {C}ommittee for {S}pace
  {D}ata {S}ystems ({CCSDS}), Washington, DC, USA, Tech. Rep. Orange Book, Sep.
  2007.

\bibitem{Johnson2003}
S.~J. Johnson and S.~R. Weller, ``A family of irregular {LDPC} codes with low
  encoding complexity,'' \emph{{IEEE} Commun. Lett.}, vol.~7, no.~2, pp.
  79--81, Feb. 2003.

\bibitem{Xia2005}
T.~Xia and B.~Xia, ``Quasi-cyclic codes from extended difference families,'' in
  \emph{Proc. {IEEE} Wireless Commun. and Networking Conf.}, vol.~2, New
  Orleans, LA, Mar. 2005, pp. 1036--1040.

\bibitem{MacWilliams71}
F.~MacWilliams, ``Orthogonal circulant matrices over finite fields, and how to
  find them,'' \emph{J. Comb.~Theory Series A}, vol.~10, pp. 1--17, 1971.

\bibitem{Zhang1997phd}
Z.~Zhang, ``Construction of the orthogonal groups of $n \times n$ circulant
  matrices over finite fields,'' Ph.D. dissertation, Concordia University,
  Montreal, Quebec, Canada, Jan. 1997.

\bibitem{Bini2001}
D.~Bini, G.~M. Del~Corso, G.~Manzini, and L.~Margara, ``Inversion of circulant
  matrices over {$Z_m$},'' \emph{Math. Comp.}, vol.~70, pp. 1169--1182, 2001.

\bibitem{PARI2009}
\BIBentryALTinterwordspacing
(2009) {PARI/GP}. [Online]. Available: \url{http://pari.math.u-bordeaux.fr/}
\BIBentrySTDinterwordspacing

\bibitem{Hagenauer1996}
J.~Hagenauer, E.~Offer, and L.~Papke, ``Iterative decoding of binary block and
  convolutional codes,'' \emph{{IEEE} Trans. Inform. Theory}, vol.~42, no.~2,
  pp. 429--445, Mar. 1996.

\bibitem{Kamiya2007}
N.~Kamiya, ``High-rate quasi-cyclic low-density parity-check codes derived from
  finite affine planes,'' \emph{{IEEE} Trans. Inform. Theory}, vol.~53, no.~4,
  pp. 1444--1459, Apr. 2007.

\bibitem{DVB-RCS}
\emph{Digital Video Broadcasting {(DVB)}; Interaction channel for satellite
  distribution systems}, {ETSI EN} Std. 301 790 (v1.4.1), Sep. 2005.

\bibitem{Lestable2006}
T.~Lestable, E.~Zimmerman, M.-H. Hamon, and S.~Stiglmayr, ``Block-{LDPC} codes
  vs duo-binary turbo-codes for european next generation wireless systems,'' in
  \emph{Proc. {IEEE VTC}-2006 Fall}, Montreal, Canada, Sep. 2006, pp. 1--5.

\bibitem{Baldi2009}
M.~Baldi, G.~Cancellieri, and F.~Chiaraluce, ``Finite-precision analysis of
  demappers and decoders for {LDPC}-coded {M-QAM}-systems,'' \emph{{IEEE}
  Trans. Broadcast.}, vol.~55, no.~2, pp. 239--250, Jun. 2009.

\end{thebibliography}

\end{document}